\documentclass[twocolumn,aps,pra,showpacs,raggedbottom,nobalancelastpage,amssymb,superscriptaddress,longbibliography]{revtex4-1}

\usepackage{graphicx}
\usepackage{amsmath}
\usepackage{amssymb}
\usepackage{bm}
\usepackage[usenames]{color}
\usepackage{amsfonts}
\usepackage{braket}
\usepackage{comment}

\usepackage[colorlinks=true , citecolor=blue,urlcolor=blue]{hyperref}

\def\kb{{\mathchar'26\mkern-9mu k}}

\usepackage{ulem}

\def\kbar{{\mathchar'26\mkern-9mu k}}

\begin{document}
\title{Effective thermalization of a many-body dynamically localized Bose gas}

\author{Vincent Vuatelet}
\affiliation{Universit\'e  de  Lille,  CNRS,  UMR  8523  --  PhLAM  --  Laboratoire  de Physique  des  Lasers,  Atomes  et  Mol\'ecules,  F-59000  Lille,  France}
\author{Adam Ran\c con}
\affiliation{Universit\'e  de  Lille,  CNRS,  UMR  8523  --  PhLAM  --  Laboratoire  de Physique  des  Lasers,  Atomes  et  Mol\'ecules,  F-59000  Lille,  France}

\begin{abstract}
Dynamical localization is the analog of Anderson localization in momentum space, where the system's energy saturates and the single-particle wave-functions are exponentially localized in momentum space. In the presence of interactions, in the context of a periodically kicked Bose gas, it has been argued that dynamical localization persists.  Focusing on the Tonks (strongly interacting) regime, we show that the many-body dynamically localized phase is effectively thermal, a clear deviation from the breaking of ergodicity observed in standard many-body localized systems. We relate the effective temperature to the driving parameters, and thus quantitatively describe the loss of coherence at large distances in this phase. Contrary to the non-interacting case, the momentum distribution decays as a power-law at large momenta, characterized by an effectively thermal Tan's contact. This is a rare example where driving and many-body (dynamical) localization lead to an effectively ergodic state.
\end{abstract}

\date{\today}
\maketitle

\section{Introduction}

Anderson localization of classical and quantum waves is a universal phenomenon induced by disorder \cite{Anderson1958,Evers2008}. Whether or not it survives in the presence of interactions has been under intense scrutiny  in recent years, both theoretically and experimentally \cite{Nandkishore2015,Abanin2019}.
 It is now well understood that while interactions tend to destroy localization, a strong enough disorder will give rise to Many-Body Localization (MBL), at least in low dimensions. MBL can be understood in terms of an effective integrability due to the existence of an extensive number of local integrals of motions, breaking ergodicity and preventing thermalization \cite{Serbyn2013,Huse2014}. The same mechanism prevents driven MBL systems from absorbing an infinite amount of energy and thus prevents runaway heating \cite{Ponte2015,Ponte2015a}.
 
Dynamical localization is the quantum chaos analog of Anderson localization, but taking place in momentum space \cite{Fishman1982}. In the paradigmatic quantum kicked rotor (QKR), periodic kicks give rise to a ballistic propagation in momentum space, while the (pseudo) random phase accumulated during the free propagation in between kicks by each momentum state plays the role of disorder, resulting in destructive  quantum interferences and dynamical localization. Experimental realizations of the atomic QKR have allowed for detailed investigations of the Anderson physics: observation of Anderson transition \cite{Chabe2008}, characterization of its critical properties \cite{Lemarie2010,Lopez2012}, localization at the upper critical dimension \cite{Manai2015}, the effects symmetries on weak localization \cite{Hainaut2018b}, and classical-to-quantum transition at early times \cite{Hainaut2018a}.

Whether interactions destroy dynamical localization is a fundamental question that challenges our understanding of driven interacting quantum systems. This has been studied for various toy-models  \cite{Adachi1988,Lei2009,Keser2016,Rozenbaum2017,Notarnicola2018,Notarnicola2020}, as well as for the kicked Lieb-Liniger model, a realistic model for cold atoms experiments \cite{Bloch2008}.
 At the mean-field level, it has been argued both on theoretical and numerical grounds that interactions destroy dynamical localization, which is replaced by a subdiffusion in momentum space \cite{Shepelyansky1993,Pikovsky2008,Flach2009,Gligoric2011,Cherroret2014,Lellouch2020}. However, it is well known that mean-field theory breaks down in one dimension \cite{Cazalilla2011}, questioning these predictions.
 Beyond mean-field, an early study for two bosons hinted that interactions may also destroy dynamical localization \cite{Qin2017}, but the validity of these results has been recently questioned \cite{Chicireanu2021}. Finally, Rylands et al. have argued that dynamical localization persists in the presence of interactions, leading to a Many-Body Dynamically Localized (MBDL) phase \cite{Rylands2020}. The MBDL phase can be described by a steady-state density matrix $\hat \rho_{ss}$, which in general should belong to a generalized Gibbs ensemble \cite{Lazarides2014,Vidmar2016}. However, this  regime and its density matrix have yet to be characterized.

In this article, we study the MBDL phase of the kicked Lieb-Liniger gas in the infinite interaction (Tonks) regime. Our main result is that the steady-state of the system is very well described by the density matrix of a {\it thermal} gas, seemingly in contradiction with the fact that the system is integrable and has an extensive number of conserved charges (the occupation of the Floquet eigenstates). We stress that this effective thermalization takes place while the system is still periodically driven. This is a rare instance where driving and Many-Body (Dynamical) Localization give rise to an effectively ergodic state. We relate this temperature to the system's parameters (kicks strength and period). This allows us to quantitatively  characterize two experimentally relevant observables: the momentum distribution that does not decay exponentially, as in the non-interacting limit, but as a power-law, which is to be expected for interacting quantum systems \cite{Olshanii2003,Tan2008}; and the coherence function, which decays exponentially, demonstrating the absence of phase coherence.

The article is organized as follows. Sec.~\ref{sec_model}, we present the model and the physical observables studied. In Sec.~\ref{sec_MBDL}, we summarize our numerical results for the momentum distribution and the coherence function in the localized regime, while in Sec.~\ref{sec_thermalization} we interpret those results in terms of an effective thermalization of the system, and we give arguments to explain such a thermalization in Sec.~\ref{sec_expl}. The discussion of our results appears in Sec.~\ref{sec_concl}.

\section{Model \label{sec_model}} 
We consider $N$ interacting bosons of mass $m$, the dynamics of which is described by the periodic Hamiltonian
\begin{equation}
\hat H(t) =  \sum_i \left(\frac{\hat p_i^2}2+K\cos(\hat x_i)\sum_n \delta(t-n)\right)+g\sum_{i<j}\delta(\hat x_i-\hat x_j).
\end{equation}
The one-body term corresponds to the QKR Hamiltonian $\hat H_{QKR}(t)$, while the other describes the  contact interaction (we also define $\hat H_{TG}=\hat H|_{K=0}$). Here and in the following, time is in units of the period $\tau$ of the kicks and length in unit of the inverse of the kick-potential wavenumber $k_K$. Momenta are normalized such that $[\hat x_i,\hat p_j]=i\kbar \delta_{ij}$, with $\kbar=\hbar k_K^2\tau/m$ the effective Planck constant. The system is of size $L=2\pi$, and we assume periodic boundary conditions, implying that momenta are quantized in unit of $\kbar$ (we will use units such that the Boltzmann constant $k_B=1$).

In the free case $(g=0)$, we recover the physics of the QKR, where any single-particle wave-function is localized in momentum space at long time (larger than the localization time) and decays exponentially  in momentum space, with the same ``localization length'' $p_{loc}$ (for larger $K/\kbar$, one finds $p_{loc}\propto K^2/\kbar$ \cite{Shepelyansky1986,Lemarie2009})
\footnote{To be precise, we define $p_{loc}$ of a single-particle state  \unexpanded{$ | \psi \rangle$} as \unexpanded{$p_{loc}^2=\langle\psi|\hat p^2|\psi\rangle-\langle\psi|\hat p|\psi\rangle^2$}, which is proportional to the momentum scale on which the dynamically localized states decay exponentially.}. In particular, the total energy of the system saturates to a constant value at long time.

Here, we focus on the Tonks regime, $g\to\infty$, allowing us to write the exact time-dependent wave-function $\Psi_B(\{x\};t)$ of the system using the Bose-Fermi mapping \cite{Girardeau1960,Lenard1964,Buljan2008,Jukic2008,Pezer2009},
\begin{equation}
\Psi_B(\{x\};t)=\prod_{i<j}{\rm sign}(x_i-x_j)\Psi_F(\{x\};t),
\end{equation}
 where $\Psi_F(\{x\};t)=\frac{1}{\sqrt{N!}}\det[\psi_i(x_j,t)]$ is the free fermions wave-function constructed from the $N$ single-particle orbitals $\psi_i(x,t)$, which evolve according to the QKR Hamiltonian, $i\kbar \partial_t |\psi_i(t)\rangle=\hat H_{QKR}(t)|\psi_i(t)\rangle$. 
We assume that the system starts in its ground state, i.e. the fermionic wave-function describes a Fermi sea with Fermi momentum $p_F\propto N$ and ground state energy $E_0$.

\begin{figure}[t]
	\includegraphics[scale=0.2,clip]{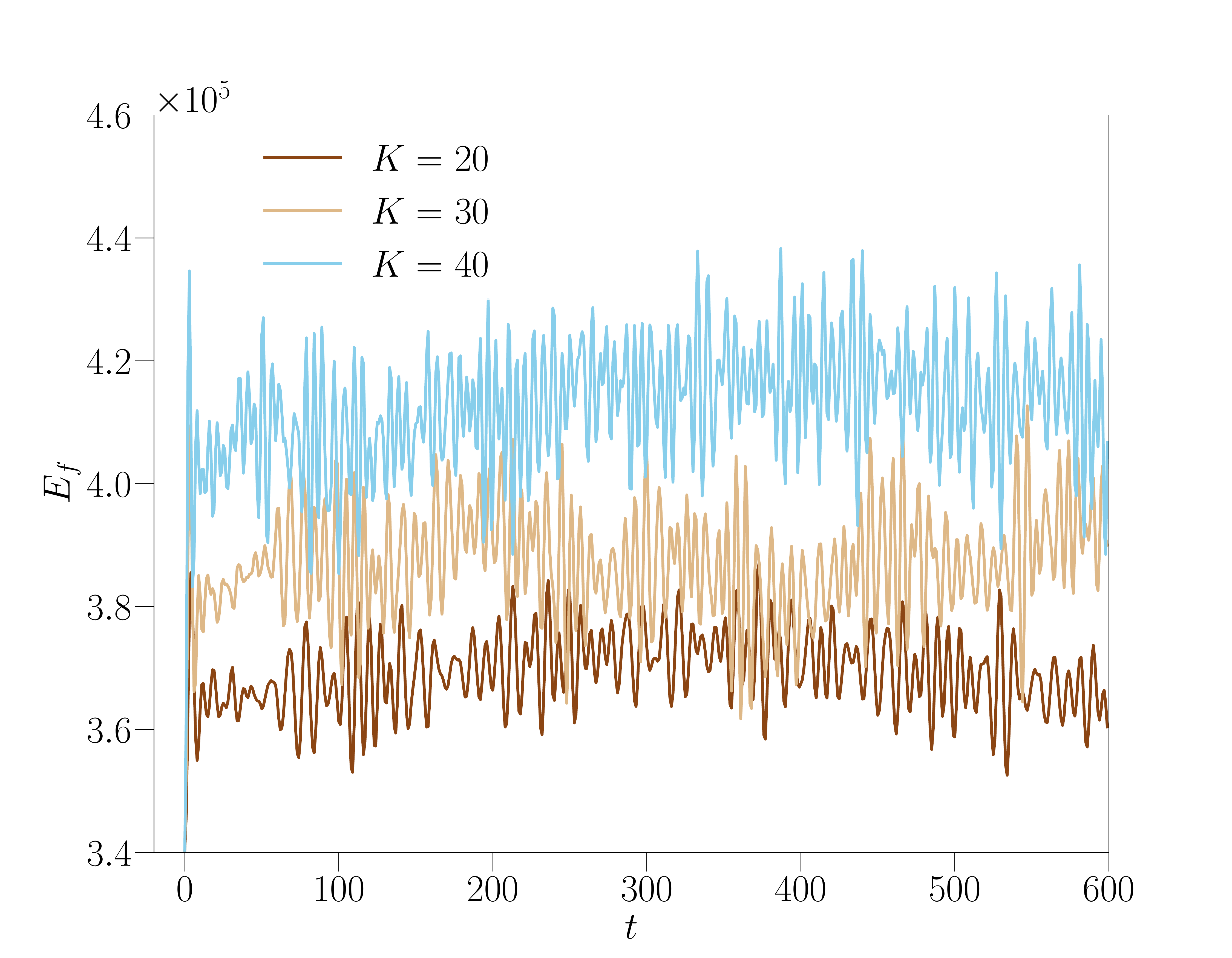}
	\caption{Time evolution of the total energy of the system $E_{f}(t)$ for $N=61$ particles for $K =  20, 30$ and $40$ ($\kb=6$).}
	\label{energy_saturation}
\end{figure}

For a Tonks gas, all bosonic local observables (such as the energy or the density) are given by those of free fermions. Therefore, since the dynamics of the single-particle orbitals $\psi_i(x,t)$ is that of the non-interacting QKR, we directly infer that they all dynamically localize at long time. The energy (of both fermions and bosons) will thus saturate to a finite value $E_f\simeq E_0+N \frac{p_{loc}^2}{2}$ for time larger than the localization time, which is interpreted as MBDL  \cite{Rylands2020}, see Fig.~\ref{energy_saturation}.
Since the fermions orbitals reach a steady-state in the MBDL phase, we expect the system to be described by a steady-state density matrix $\hat \rho_{ss}$, belonging a priori to the generalized Gibbs ensemble \cite{Lazarides2014}, see discussion in Sec.~\ref{sec_expl}. Here we focus on the properties of the system in this MDBL steady state, and thus do not write time dependence of observables.

Non-local observables such as the steady-state one-body density matrix (OBDM) 
\begin{equation}
\begin{split}
\rho(x,y) = N\int d x_2\ldots dx_N &\Psi_B^*(x,x_2,\ldots,x_N)\times\\
&\Psi_B(y,x_2,\ldots,x_N),
\end{split}
\end{equation}  
and its Fourier transform, the momentum distribution 
\begin{equation}
n_k=\frac1L\int dx dy\, e^{ik(x-y)}\rho(x,y),
\end{equation}  
are significantly different with those of free fermions. Since dynamical localization is a non-local phenomenon, we therefore expect these observables to significantly differ from that of free particles \footnote{This is to be contrasted with the case of a Tonks gas loaded in a random potential, where localization in real space of the fermions is naturally preserved for the bosons \cite{Radic2010,Seiringer2016}.}. We therefore focus on those observables in the steady-state (time much larger than the localization time) in the following.

The time-evolution of each single-particle orbital is performed numerically by discretizing space and using Fast Fourier Transform to alternate between real space for the kicks and momentum space for the free propagation. 
The observables are computed using the method of Refs.~\cite{Rigol2005a,Rigol2005}.

\begin{figure}[t]
 \includegraphics[scale=0.2,clip]{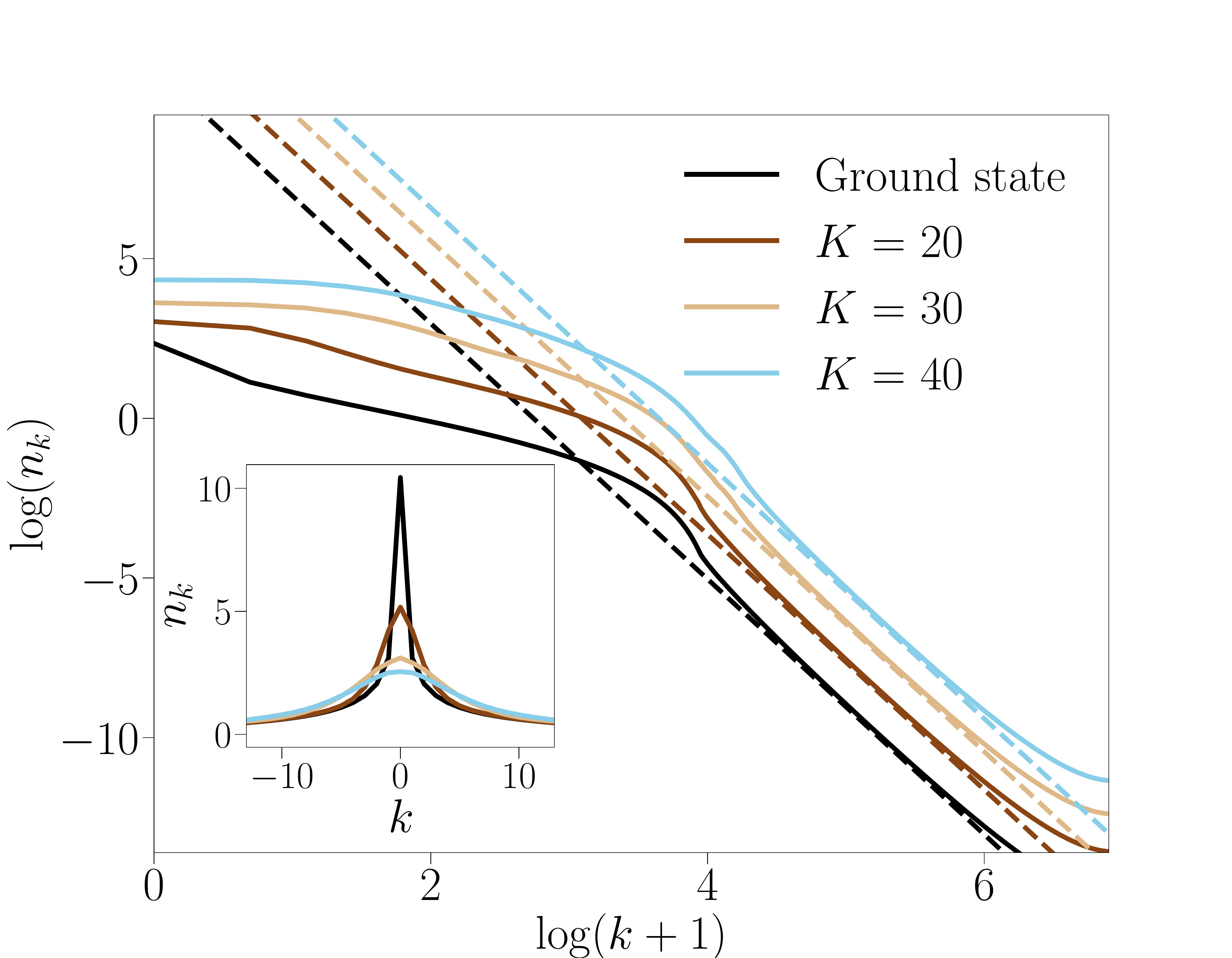}
\caption{Steady-state momentum distribution for $N=51$ particles at $\kbar=6$ for $K=20$, $30$ and $40$ in log-log scale (the different $n_k$ have been shifted for better visibility in the main panel).
The dashed line shows the asymptotic behavior $n_k\simeq \mathcal C_{ss}/k^{4}$ at large momenta, with $\mathcal C_{ss}$ computed using the effectively thermal density matrix (see text).The inset shows the same quantities in linear scale.}
\label{fig_nk}
\end{figure}

\begin{figure}[t]
\includegraphics[scale=0.2,clip]{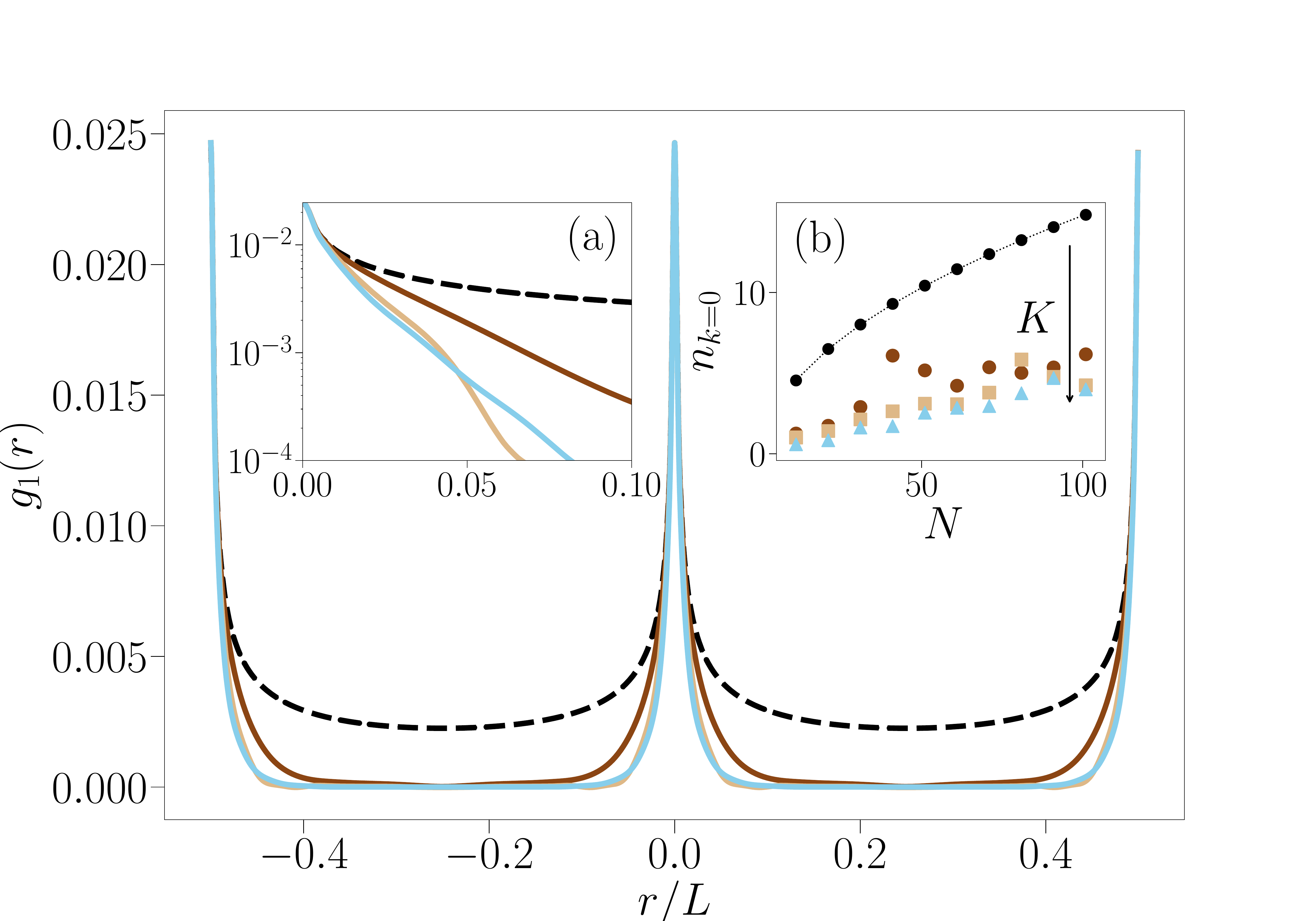}
\caption{Steady-state coherence function $g_1(r)$ for $N=101$ particles at $\kbar=6$ for $K=20$, $30$ and $40$. Insets (a): Same data, in semi-log scale, emphasizing the exponential decay in the MBDL, compared to the $1/\sqrt{r}$ decay of the initial condition (dashed curve); (b): Occupation of the zero-momentum state $n_{k=0}$.  It grows as $\sqrt{N}$ in the ground state (dotted line), but saturate to a finite value in the MBDL regime.}
\label{fig_g1}
\end{figure}

\section{MBDL momentum distribution and coherence \label{sec_MBDL}} 
 The groundstate of the Tonks gas is characterized by quasi-long-range order, $n_k\propto 1/\sqrt{k}$ at small momenta and $n_{k=0}\propto \sqrt{N}$, where the sublinear scaling implies the absence of true long-range order \cite{Vaidya1979}.  Fig. \ref{fig_nk}  shows the  momentum distribution in the ground state and in the localized regime for  $N=51$ bosons, $\kbar=6$ and various values of $K$, in log-log scale. The divergence at small momenta of the momentum distribution is rounded (see inset), while we observe  a power-law decay at large momenta, $n_k\simeq \mathcal C/k^{4}$.
  This behavior is a universal feature of interacting quantum systems, where $\mathcal C$ is the so-called Tan's contact \cite{Olshanii2003,Tan2008}. We conclude that while the interactions do not destroy dynamical localization, in the sense that the system does not heat up to infinite temperature, they do significantly alter the exponential localization in momentum distribution of the bosons.
 
The coherence of the Tonks gas in the MBDL regime can also be characterized by the coherence function 
\begin{equation}
g_1(r)=\frac{1}{L}\int dR\,  \rho(R-r/2,R+r/2).
\end{equation}
 In its ground state, the gas has algebraic correlations, $g_1(r,t=0)\propto 1/\sqrt{r}$, corresponding to quasi-long-range order \cite{Cazalilla2011}. Fig.~\ref{fig_g1} shows that in the MBDL regime, the coherence function decays exponentially fast at large distance, implying that the kicks have destroyed the coherence of the quasi-condensate. This is in agreement with the fact that $n_{k=0}$ does not scale with the number of particles (see inset (b) of Fig.~\ref{fig_g1}).

\section{Effective thermalization of MBDL \label{sec_thermalization}} 
	
The absence of quasi-long-range coherence of the localized regime is similar to that of a thermal Tonks gas \cite{Cazalilla2011}. We now show that the system is very well described in the MDBL by the a thermal density matrix $\hat \rho_{ss}\simeq \hat \rho_{th}$, where $\hat \rho_{th}$ is the thermal density matrix of the Tonks gas
\begin{equation}
 \hat \rho_{th}\propto e^{-(\hat H_{TG}-\mu_{eff} \hat N)/T_{eff}},
 \end{equation} 
with effective temperature $T_{eff}$ and effective chemical potential $\mu_{eff}$ that depends on the system's parameters and  the number of particles. 

Thanks to the Bose-Fermi mapping, if there is indeed effective thermalization, we expect the momentum distribution $n_k^F$ of the underlying free fermions to be described by a Fermi-Dirac distribution, allowing us to extract $T_{eff}$ and $\mu_{eff}$. We therefore begin by analyzing the thermal properties of the free fermions, and then address the thermal-like properties of the Tonks gas in the localized regime.

\subsection{Fermions}
	
\begin{figure}[t!]
	\begin{center}
	\includegraphics[scale=0.2,clip]{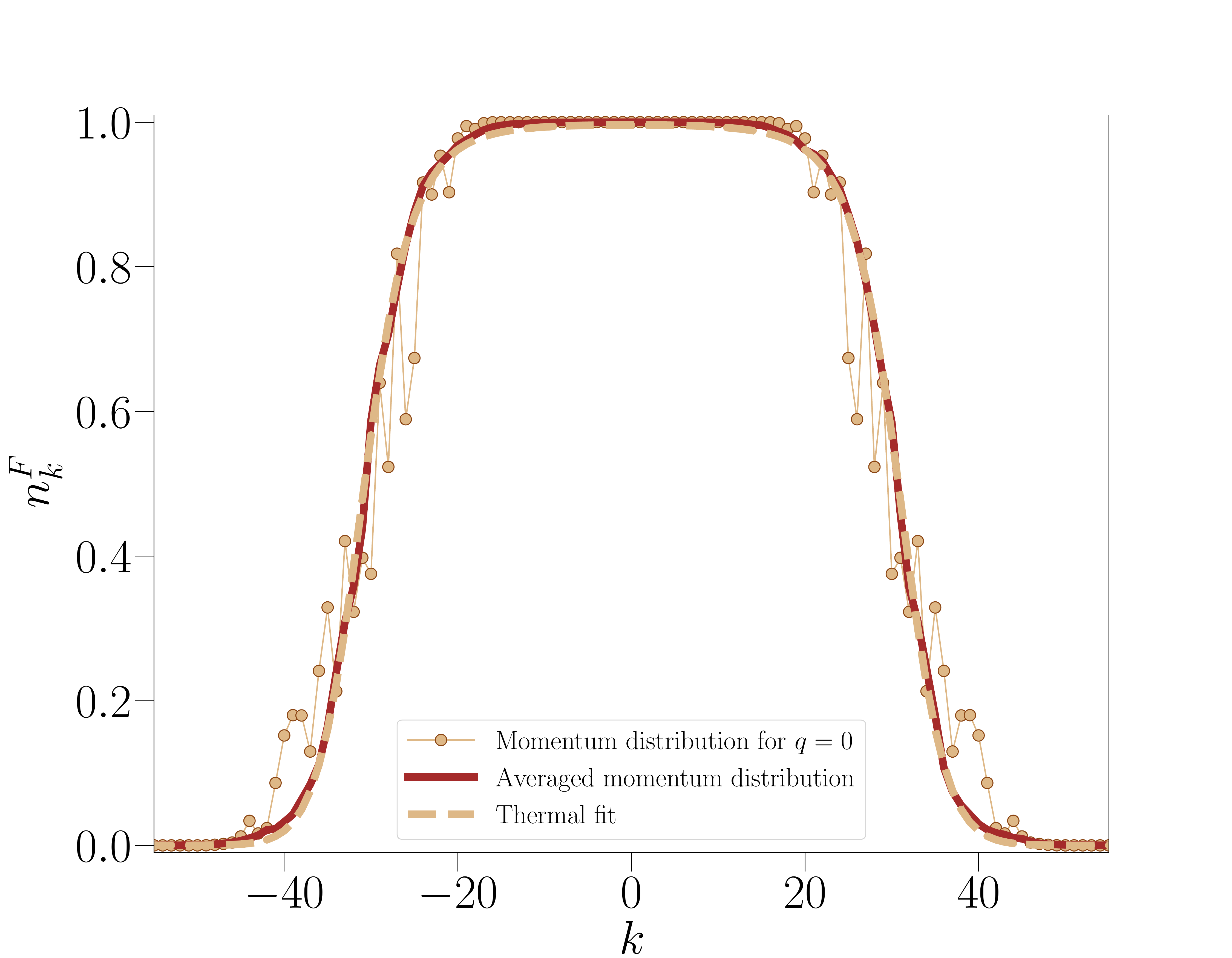}
	\end{center}
	\caption{Comparaison between raw and averaged distribution for the many-body momentum distribution. In this case, $N=61$, $K=30$, $\kbar=6$. \label{fig comparaison raw average}}
\end{figure}

An example of the momentum distribution of the fermions in the localized regime is shown in Fig.~\ref{fig comparaison raw average} (symbols). Contrary to the momentum distribution of the bosons, it is rather noisy, as typical for disordered systems. To better fit the momentum distribution of the fermions, it is convenient to introduce a modified QKR Hamiltonian depending on a parameter $q$ \footnote{The quantity $q$ plays the role of a conserved quasi-momentum in the single-particle QKR without periodic boundary conditions.}
\begin{equation}
\hat{H}_{q}=\frac{(\hat{p}+q\kbar)^{2}}{2}+K\cos(\hat{x})\sum_{n}\delta(t-n).
\end{equation} 
Note that we never average the bosonic observables (we always consider the physical value $q=0$), such as the OBDM or the momentum distribution. The highly non-linear transformation relating the bosonic observables to that of the fermions   averages out the fluctuations. In App.~\ref{app_fit}, we show below that the temperature that can be estimated from $q=0$  is very well correlated with that extracted from the average fermionic distribution.

In Fig.~\ref{fig comparaison raw average}, in addition to the momentum distribution $n_{k}^F$ at $q=0$ discussed above, we also show the $n_{k}^F$ the momentum distribution averaged over 150 random values of $q$ (full line). The smoothing effect of the averaging procedure is very clear. On the same figure, we also show a Fermi-Dirac distribution at an effective temperature $T_{eff}$ and effective chemical potential $\mu_{eff}$ such that this thermal distribution explains very well the data (dashed line).

\begin{figure}[t]
	\includegraphics[scale=0.2,clip]{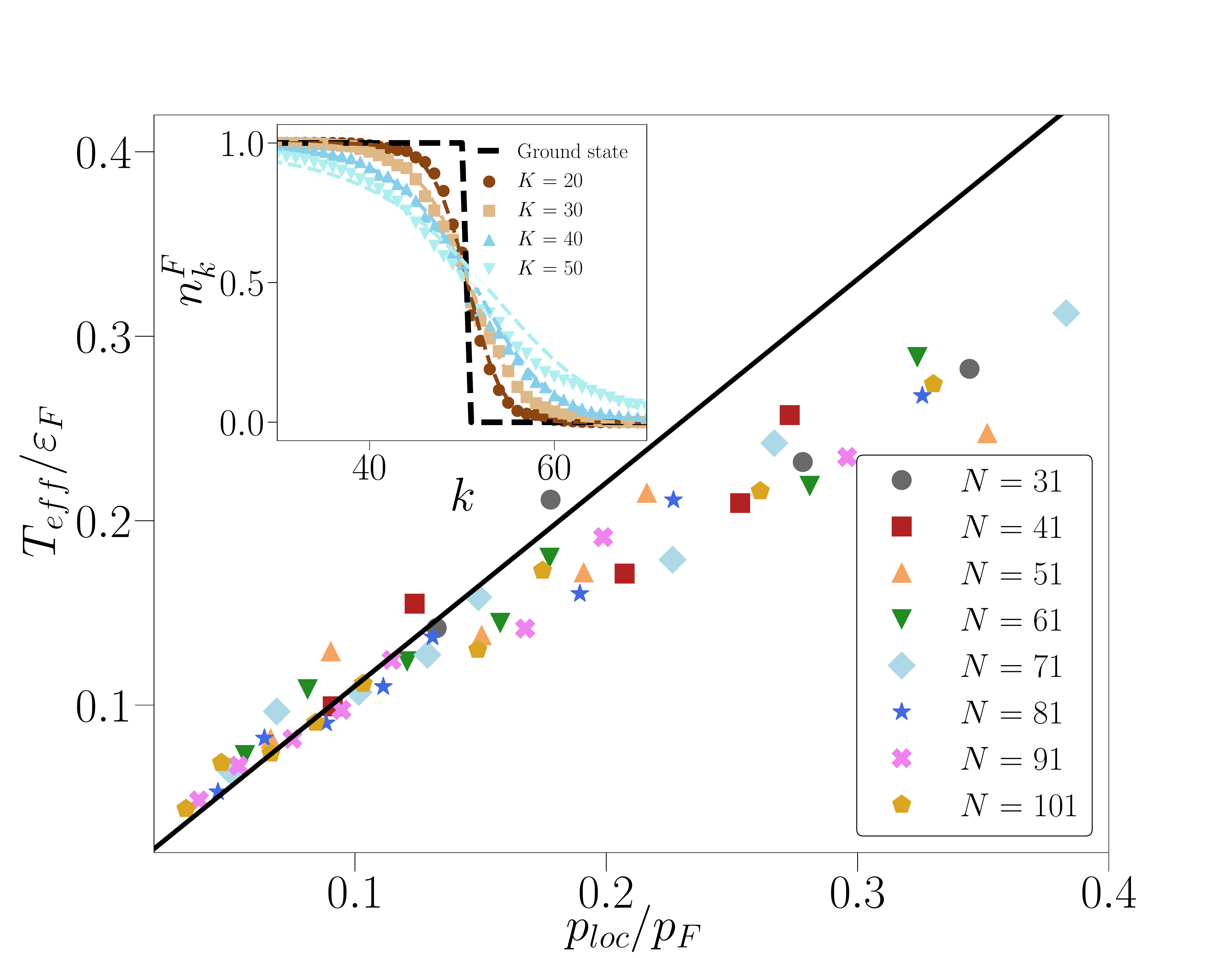}
	\caption{Effective temperature $T_{eff}/\varepsilon_F$ as a function of $p_{loc}/p_F$ for various particle numbers. The collapse of the data shows the linear scaling for small enough $p_{loc}/p_F$, $T_{eff}/\varepsilon_F\simeq  \frac{2\sqrt{3}}\pi p_{loc}/p_F$ (black line).  Inset: Momentum distribution of the fermions $n_k^F$ in the localized regime (symbols), fitted by a Fermi-Dirac distribution with temperature $T_{eff}$ and chemical potential $\mu_{eff}$, for $N=101$ and $\kbar=6$.}
	\label{fig_fermion_thermal}
\end{figure}
The effective temperature and chemical potential are obtained by imposing that 
\begin{equation}
\begin{split}
&\sum_{k\in \mathbb Z}f_{FD}(k,T_{eff},\mu_{eff})=N,\\
&\sum_{k\in \mathbb Z} \frac{\kbar^2k^2}2 f_{FD}(k,T_{eff},\mu_{eff})=E_{f},
\end{split}
\end{equation}
where $E_f$ is the energy obtained from the averaged momentum distribution $n_{k}^F$, and $f_{FD}$ is the Fermi-Dirac distribution
\begin{equation}
f_{FD}(k,T,\mu)=\frac{1}{e^{\frac{\kbar^2k^2-\mu}{2T}}+1}.
\end{equation}
We observe that  the fit is very good, see the inset of Fig.~\ref{fig_fermion_thermal}, as long as $p_F\gg p_{loc}$  (corresponding to small enough $K$), see also App.~\ref{app_fit} for a detailed analysis of the parameters regime where the thermal fit works. We focus on this effectively thermal regime here. This corresponds to low effective temperatures compared to the initial Fermi energy $\varepsilon_F=p_F^2/2$, which allows us to find an explicit expression of the effective temperature in terms of the two natural quantities $p_{loc}$ and $p_F$.

The initial condition of the system corresponds to the ground state, the energy of which is 
\begin{equation}
E_0=\frac{N \varepsilon_F}{3},
\end{equation}
for a one-dimensional Fermi gas, with $\varepsilon_F=\frac{p_F^2}{2}$ the Fermi energy, which in our units read $\varepsilon_F=\frac{N^2}{8}$ ($N\gg 1$).
On the other hand, in the localized regime, the final energy reads
\begin{equation}
E_{f} =  E_{0}+N\frac{p_{loc}^{2}}{2}.
\end{equation}
Assuming that the system is thermal, the Sommerfeld expansion of the energy gives
\begin{equation}
E(T_{eff})\simeq \frac{N\varepsilon_F}{3}+\frac{N\pi^2}{12}\frac{T_{eff}^2}{\varepsilon_F}+\ldots,
\label{eq_energy}
\end{equation}
Equating $E_{f}=E(T_{eff})$, we obtain
\begin{equation}
\frac{T_{eff}}{\varepsilon_F}=\frac{2\sqrt{3}}\pi \frac{p_{loc}}{p_F}.
\label{eq_Teff_ploc}
\end{equation}
Note that the effective temperature is indeed small (compared to the Fermi energy) for small $p_{loc}/{p_F}$, validating our initial assumption.

 Fig.~\ref{fig_fermion_thermal} shows that indeed Eq.~\eqref{eq_Teff_ploc} works very well for $p_{loc}/{p_F}\ll 1$. Note that while the effective temperature scales linearly with the particle number, the relative thermal broadening of the Fermi distribution $T_{eff}/\varepsilon_F$ vanishes as $N^{-1}$.


\subsection{Implications for the bosons}

Assuming that the steady-state density matrix  $\hat \rho_{ss}$ is thermal allows us to quantitatively characterize the momentum distribution and the coherence function  of the Tonks gas in the localized regime, an a priori formidable task without this insight. 

At short distance, the coherence function of a Tonks gas is known to be non-analytic due to the interactions, $g_1(r)\sim \frac{\pi\mathcal C}{6 L}|r|^3$. For a thermal Tonks gas of $N$ bosons at temperature $T$, the contact reads $\mathcal C_{th}(T,N)=\frac{8 NE(T,N)}{ L^2 \kbar^2}$ \cite{Vignolo2013}. We therefore infer that the contact in the MBDL regime $\mathcal C_{ss}$ should be given by  $\mathcal C_{ss}=\mathcal C_{th}(T_{eff},N)$.  Fig.~\ref{fig_nk} shows that the power-law decay is very well explained by $\mathcal C_{th}(T_{eff},N)/k^4$, showed as dashed lines.

\begin{figure}[t]
\includegraphics[scale=0.2,clip]{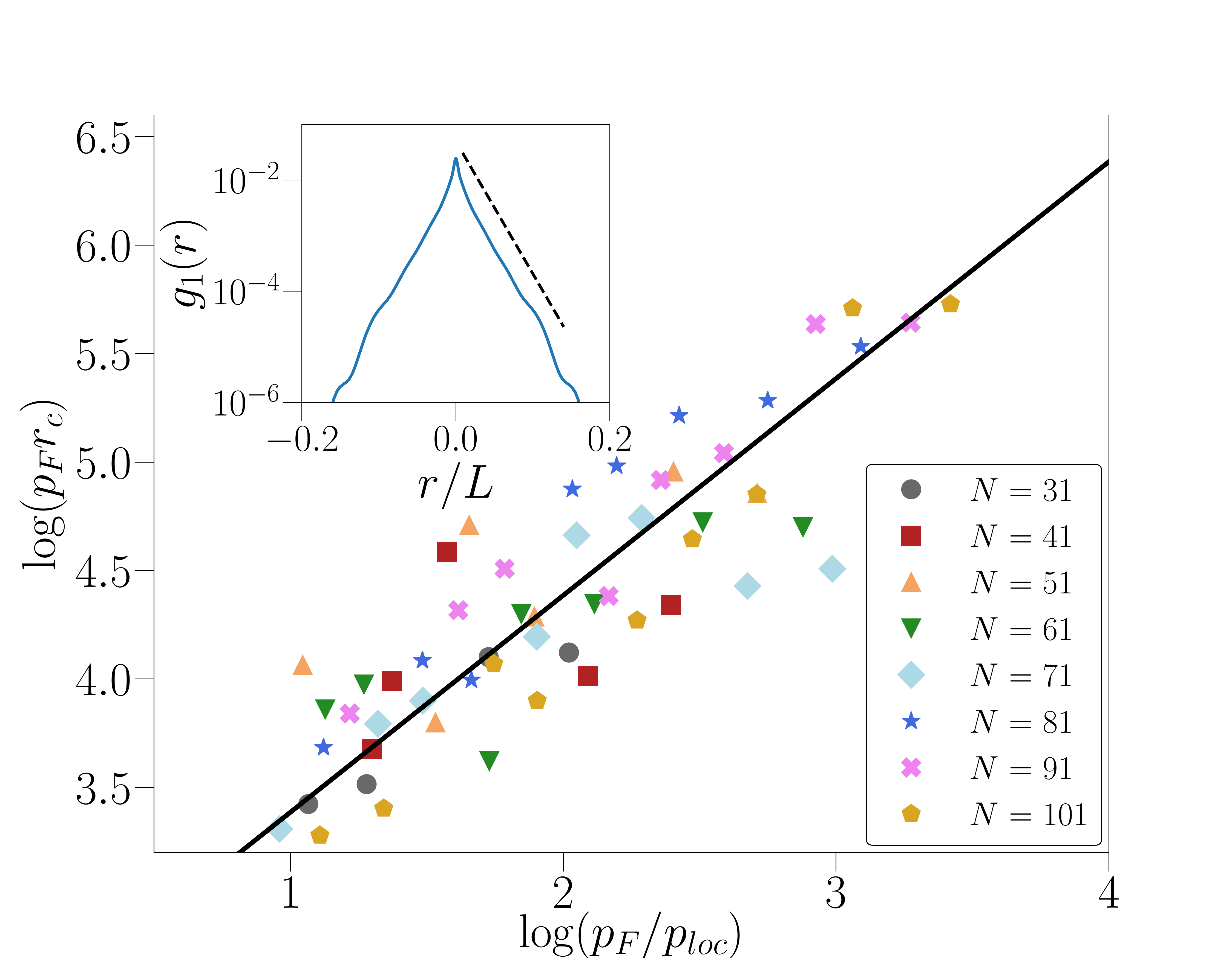}
\caption{Correlation length $r_c$ as a function of $p_{loc}$ for various $N$. The collapse of the data shows that it is independent of the particle number.  The line corresponds to the scaling $r_c=\frac{\pi}{\sqrt{3}}\frac{\kbar}{p_{loc}}$. Inset: Coherence function $g_{1}(r)$ (blue line) and the expected exponential decay with $r_c=\frac{\kbar\pi}{\sqrt{3}p_{loc}}$, for $N=101$, $K=40$, $\kbar=6$. }
\label{fig_xi}
\end{figure}

 At long distances, the exponential decay of $ g_1(r)$ of a Tonks gas at finite temperature, $g_1(r)\propto e^{-2|r|/r_c}$, is also known \cite{Cazalilla2011,Its1991}, and in the low-temperature limit we expect $r_c=\frac{\kbar v_F}{T_{eff}}$, where $v_F=\frac{\kbar N}{2}$ is the Fermi velocity in our units. Therefore, due to the effective thermality of the MBDL phase, we expect the correlation length $r_c$   to be independent of the particle number and to be inversely proportional to $p_{loc}$. Combining the thermal scaling and the expression in Eq.~\eqref{eq_Teff_ploc}, we do expect the scaling $r_c=\frac{\pi}{\sqrt{3}}\frac{\kbar}{p_{loc}}$, which can be rewritten as
\begin{equation}
r_{c}p_{F}=\frac{\kbar\pi}{\sqrt{3}}\frac{p_{F}}{p_{loc}}.
\label{eq_rc}
\end{equation}
Fig.~\ref{fig_xi} shows good agreement between Eq.~\eqref{eq_rc} and the correlation length extracted from the steady-state (see App.~\ref{app_rc} for details). 
The inset shows that Eq.~\eqref{eq_rc} describes well the exponential decay of the coherence function. 

\section{Explanation of the effective thermalization \label{sec_expl}} 
\begin{figure}[t]
	\includegraphics[scale=0.2,clip]{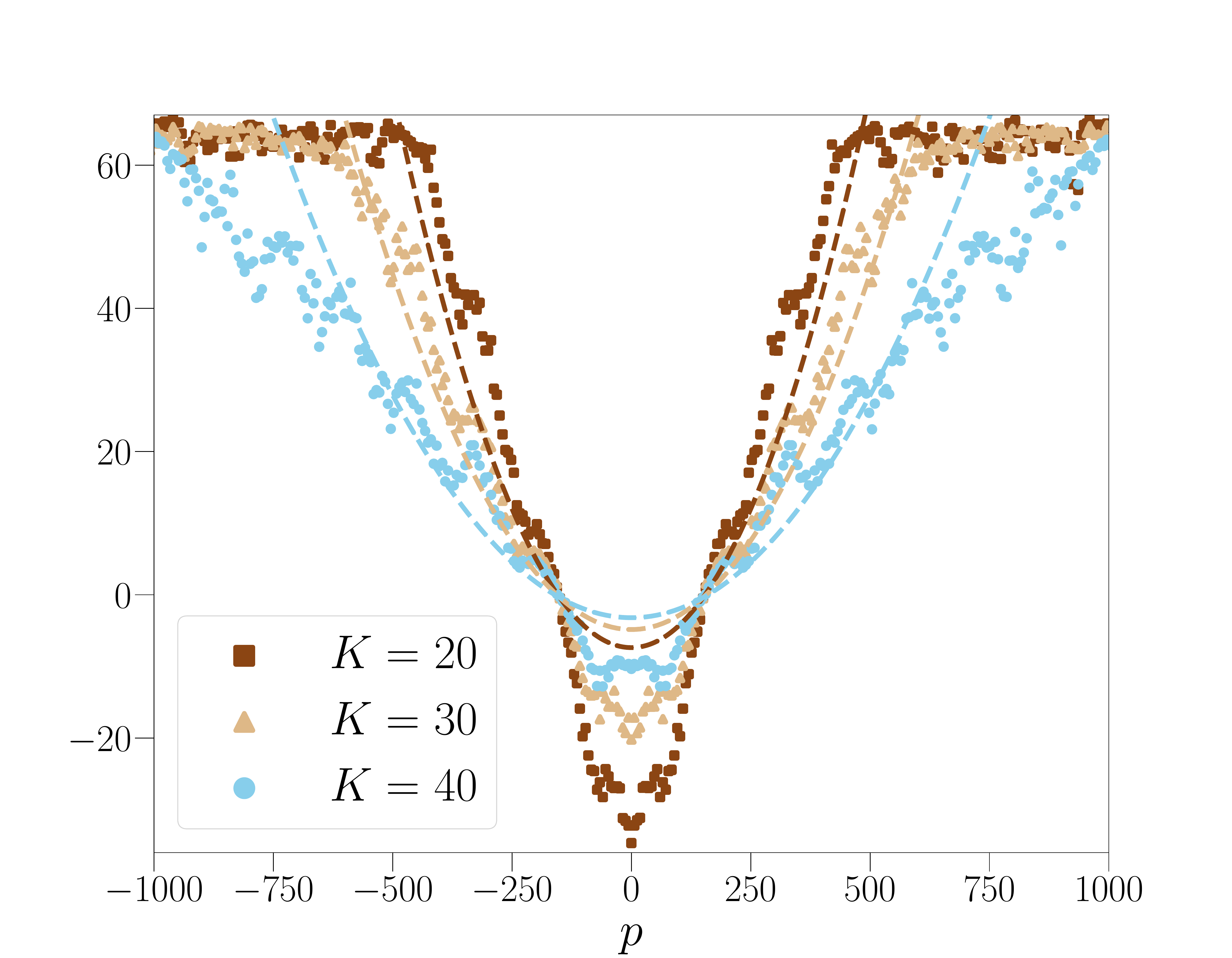}
	\caption{Comparison of $h_{p}$ (symbols) with $f_{p}$ for varying $K$, $N=51$ and $\kb=6$.}
	\label{hp compared to thermal approx}
\end{figure}

Let us now argue why the MBDL steady-state appears thermal. This is best understood using the fermionic degrees of freedom, which are non-interacting and evolve according to $\hat H_{QKR}$. Introducing the evolution operator over one period $\hat U_{QKR}$ and its Floquet eigenstates $\hat U_{QKR}|\phi_\alpha\rangle=e^{-i\omega_\alpha}|\phi_\alpha\rangle$, it can be written in second quantization as $\hat U_{QKR}=\exp\left(-i\sum_\alpha \omega_\alpha \hat f^\dag_\alpha \hat f_\alpha\right)$. Now, the occupation numbers of the Floquet eigenstates $n_\alpha=\langle \hat f^\dag_\alpha \hat f_\alpha\rangle$  are obviously constants of motion, and since there is an extensive number of them, the system is integrable. We therefore expect the steady-state to be described by the (periodic) generalized Gibbs  ensemble (GGE) \cite{Collura2013,Lazarides2014}, $\hat \rho_{ss} \simeq  \hat \rho_{GGE}$, with
\begin{equation}
 \hat \rho_{GGE}\propto e^{-\sum_\alpha \lambda_\alpha  \hat f^\dag_\alpha \hat f_\alpha},
 \end{equation} 
where the Lagrange multipliers $\lambda_\alpha=\log((1-n_\alpha)/n_\alpha)$ are such that ${\rm Tr}\left( \hat \rho_{GGE} \hat f^\dag_\alpha \hat f_\alpha\right)=n_\alpha$ 
\footnote{The Periodic Gibbs ensemble allows for a time dependence of the steady-state density matrix $\rho_{GGE}(t)$ with $0\leq t<1$, with period $1$. We only consider here the case $t=1^-$ (just before the next kick). We have checked that other choices do not change the qualitative picture.}.
It is a priori surprising that this GGE density matrix, depending on an extensive number of Lagrange multipliers, is well described by a thermal density matrix, depending only on $T_{eff}$ and $\mu_{eff}$. To understand this, we write it 
in terms of a non-local operator in momentum space
\begin{equation}
 \hat \rho_{GGE}\propto e^{-\sum_{p,q}M_{p,q}  \hat f^\dag_p \hat f_{q}},
 \end{equation} 
with $M_{p,q} =\sum_{\alpha}\langle p|\phi_\alpha \rangle \lambda_{\alpha}\langle\phi_\alpha| q\rangle$.  Therefore, for generic dynamics and initial states, one should expect a large number of non-local conserved quantities and a clear departure from a thermal state.
However, in the present case, we note that the Floquet eigenstates are exponentially localized in momentum space, over a scale of order $p_{loc}$\cite{Fishman1982}, implying that: (i)  only the states with $|p_\alpha|\lesssim p_F+p_{loc}$ are occupied ($n_\alpha\simeq 1\,({\rm resp.}\,0)$ for $|p_\alpha|\ll p_F$ (resp. $|p_\alpha|\gg p_F$)), with $n_\alpha$ interpolating between $1$ and $0$ around $|p_\alpha|\simeq p_F$ on a width of order $p_{loc}$; (ii) $M_{p,q}\simeq 0$ if $|p-q|\gg p_{loc}$, meaning that it is almost diagonal, $M_{p,q}\simeq \delta_{p,q}h_p$ for some $h_p$. In practice, we find that $h_p\simeq f_{p}\equiv (-\mu_{eff}+p^2/2)/T_{eff}$ to a good approximation as shown in Fig.~\ref{hp compared to thermal approx}, justifying the effective thermalization $\hat \rho_{ss}\simeq \hat \rho_{th}$. Note that this effective thermalization depends crucially on point (i), as other initial conditions far from the ground state, or a too large $p_{loc}$ implying that too many eigenstate are populated, do not allow for a description of the steady-state in terms of a thermal density matrix \cite{Vuatelet_tobepublished}. 

The fact that $M_{p,q}$ is not exactly diagonal means that the steady state is not perfectly described by a thermal density matrix. In particular, it implies that the natural orbitals of the OBDM are not exactly plane-waves, but have  width $p_{loc}$ and that the two-dimensional Fourier transform of the OBDM, $L^{-1}\int dx dy\, e^{ik(x+y)}\rho(x,y;t)$ decays exponentially as $\exp(-|k|/p_{loc})$ instead of being $N\delta_{k,0}$, see App.~\ref{app_orbitals}.

\section{Conclusion \label{sec_concl}}
 We have studied the steady-state of a kicked Tonks gas. While dynamical localization is preserved by the interactions, in the sense that the system does not heat up to infinite temperature, we have shown that the momentum distribution of the bosons is not exponentially localized, as in the non-interacting case. Instead, it decays as a power-law given by Tan's contact, as expected for an interacting quantum many-body system. We have also shown that the steady-state is very well described by a thermal density matrix, with an effective temperature that scales linearly with both the Fermi and localization momenta. This steady-state is therefore both many-body dynamically localized and well described by a small number of constant of motions, corresponding to the particle number and the energy of the localized state, even though the dynamic is integrable, with an extensive number of conserved quantities. This is in contrast with standard MBL, where ergodicity breaking corresponds to emergent integrability and the existence of an extensive set of quasi-local integrals of motion \cite{Abanin2019}. 
MBDL should be observable in state-of-the-art cold atoms experiments  by measuring the steady-state momentum distribution using for instance the methods of Refs. \cite{Amerongen2008,Fabbri2011}.  
As long as the initial temperature is smaller than the effective temperature, effective thermalization should dominate 
\footnote{Using $T_{eff}\simeq \sqrt{\varepsilon_{loc}\varepsilon_{F}}$, with $\varepsilon_{loc}$ the localization energy per particles, and the estimates $\varepsilon_F\simeq E_R N^2$ and $\varepsilon_{loc}\simeq \alpha E_R$ with $E_R$ the recoil energy and $\alpha$ typically of order a few hundreds depending on the kicks parameters, we estimate that $T_{eff}\simeq 10 N T_R$ ($T_R=E_R/k_B$ the recoil temperature, typically of order $ 10^{-7} K$ in cold atoms experiments).}. It can be tested by measuring the momentum distribution of the underlying fermions \cite{Wilson2020,Malvania2021}, extracting the corresponding temperature, and comparing with the bosons' observables. 

In the few body-limit, it has been shown that finite or infinite interactions give a rather similar dynamical localization of the kicked Lieb-Liniger model \cite{Chicireanu2021}. An interesting question is whether this effective thermalization persists beyond the Tonks regime and allows for a quantitative description of the many-body dynamical localization at finite interactions. 

Finally, it is well-known that if the kicks strength is modulated, the (non-interacting) QKR displays a delocalization transition similar to the Anderson transition \cite{Shepelyansky1987,Casati1989}, which has been observed experimentally in the atomic QKR \cite{Chabe2008,Lemarie2009}. We therefore expect that modulating the kicks in the kicked Lieb-Liniger model will induce a phase transition from the MBDL to a new phase where the system can heat up to infinite temperature. Understanding the properties of such a delocalized phase is under progress.

\section*{Acknowledgments}
We thank R. Chicireanu, J.-C. Garreau, D. Delande, N. Cherroret, C. Tian and H. Buljan for  insightful discussions. This work was supported by Agence Nationale de la Recherche through Research Grants QRITiC I-SITE ULNE/ ANR-16-IDEX-0004 ULNE, the Labex CEMPI Grant No.ANR-11-LABX-0007-01, the Programme Investissements d'Avenir ANR-11-IDEX-0002-02, reference ANR-10-LABX-0037-NEXT and the Ministry of Higher Education and Research, Hauts-de-France Council and European Regional Development Fund (ERDF) through the Contrat de Projets \'Etat-Region (CPER Photonics for Society, P4S).

\bibliography{bib_paper_MDBL_Tonks_article}

\begin{thebibliography}{63}%
\makeatletter
\providecommand \@ifxundefined [1]{%
 \@ifx{#1\undefined}
}%
\providecommand \@ifnum [1]{%
 \ifnum #1\expandafter \@firstoftwo
 \else \expandafter \@secondoftwo
 \fi
}%
\providecommand \@ifx [1]{%
 \ifx #1\expandafter \@firstoftwo
 \else \expandafter \@secondoftwo
 \fi
}%
\providecommand \natexlab [1]{#1}%
\providecommand \enquote  [1]{``#1''}%
\providecommand \bibnamefont  [1]{#1}%
\providecommand \bibfnamefont [1]{#1}%
\providecommand \citenamefont [1]{#1}%
\providecommand \href@noop [0]{\@secondoftwo}%
\providecommand \href [0]{\begingroup \@sanitize@url \@href}%
\providecommand \@href[1]{\@@startlink{#1}\@@href}%
\providecommand \@@href[1]{\endgroup#1\@@endlink}%
\providecommand \@sanitize@url [0]{\catcode `\\12\catcode `\$12\catcode
  `\&12\catcode `\#12\catcode `\^12\catcode `\_12\catcode `\%12\relax}%
\providecommand \@@startlink[1]{}%
\providecommand \@@endlink[0]{}%
\providecommand \url  [0]{\begingroup\@sanitize@url \@url }%
\providecommand \@url [1]{\endgroup\@href {#1}{\urlprefix }}%
\providecommand \urlprefix  [0]{URL }%
\providecommand \Eprint [0]{\href }%
\providecommand \doibase [0]{http://dx.doi.org/}%
\providecommand \selectlanguage [0]{\@gobble}%
\providecommand \bibinfo  [0]{\@secondoftwo}%
\providecommand \bibfield  [0]{\@secondoftwo}%
\providecommand \translation [1]{[#1]}%
\providecommand \BibitemOpen [0]{}%
\providecommand \bibitemStop [0]{}%
\providecommand \bibitemNoStop [0]{.\EOS\space}%
\providecommand \EOS [0]{\spacefactor3000\relax}%
\providecommand \BibitemShut  [1]{\csname bibitem#1\endcsname}%
\let\auto@bib@innerbib\@empty
\bibitem [{\citenamefont {Anderson}(1958)}]{Anderson1958}%
  \BibitemOpen
  \bibfield  {author} {\bibinfo {author} {\bibfnamefont {P.~W.}\ \bibnamefont
  {Anderson}},\ }\bibfield  {title} {\enquote {\bibinfo {title} {Absence of
  diffusion in certain random lattices},}\ }\href {\doibase
  10.1103/PhysRev.109.1492} {\bibfield  {journal} {\bibinfo  {journal} {Phys.
  Rev.}\ }\textbf {\bibinfo {volume} {109}},\ \bibinfo {pages} {1492--1505}
  (\bibinfo {year} {1958})}\BibitemShut {NoStop}%
\bibitem [{\citenamefont {Evers}\ and\ \citenamefont
  {Mirlin}(2008)}]{Evers2008}%
  \BibitemOpen
  \bibfield  {author} {\bibinfo {author} {\bibfnamefont {Ferdinand}\
  \bibnamefont {Evers}}\ and\ \bibinfo {author} {\bibfnamefont {Alexander~D.}\
  \bibnamefont {Mirlin}},\ }\bibfield  {title} {\enquote {\bibinfo {title}
  {Anderson transitions},}\ }\href {\doibase 10.1103/RevModPhys.80.1355}
  {\bibfield  {journal} {\bibinfo  {journal} {Rev. Mod. Phys.}\ }\textbf
  {\bibinfo {volume} {80}},\ \bibinfo {pages} {1355--1417} (\bibinfo {year}
  {2008})}\BibitemShut {NoStop}%
\bibitem [{\citenamefont {Nandkishore}\ and\ \citenamefont
  {Huse}(2015)}]{Nandkishore2015}%
  \BibitemOpen
  \bibfield  {author} {\bibinfo {author} {\bibfnamefont {Rahul}\ \bibnamefont
  {Nandkishore}}\ and\ \bibinfo {author} {\bibfnamefont {David~A.}\
  \bibnamefont {Huse}},\ }\bibfield  {title} {\enquote {\bibinfo {title}
  {Many-body localization and thermalization in quantum statistical
  mechanics},}\ }\href {\doibase 10.1146/annurev-conmatphys-031214-014726}
  {\bibfield  {journal} {\bibinfo  {journal} {Annual Review of Condensed Matter
  Physics}\ }\textbf {\bibinfo {volume} {6}},\ \bibinfo {pages} {15--38}
  (\bibinfo {year} {2015})}\BibitemShut {NoStop}%
\bibitem [{\citenamefont {Abanin}\ \emph {et~al.}(2019)\citenamefont {Abanin},
  \citenamefont {Altman}, \citenamefont {Bloch},\ and\ \citenamefont
  {Serbyn}}]{Abanin2019}%
  \BibitemOpen
  \bibfield  {author} {\bibinfo {author} {\bibfnamefont {Dmitry~A.}\
  \bibnamefont {Abanin}}, \bibinfo {author} {\bibfnamefont {Ehud}\ \bibnamefont
  {Altman}}, \bibinfo {author} {\bibfnamefont {Immanuel}\ \bibnamefont
  {Bloch}}, \ and\ \bibinfo {author} {\bibfnamefont {Maksym}\ \bibnamefont
  {Serbyn}},\ }\bibfield  {title} {\enquote {\bibinfo {title} {Colloquium:
  Many-body localization, thermalization, and entanglement},}\ }\href {\doibase
  10.1103/RevModPhys.91.021001} {\bibfield  {journal} {\bibinfo  {journal}
  {Rev. Mod. Phys.}\ }\textbf {\bibinfo {volume} {91}},\ \bibinfo {pages}
  {021001} (\bibinfo {year} {2019})}\BibitemShut {NoStop}%
\bibitem [{\citenamefont {Serbyn}\ \emph {et~al.}(2013)\citenamefont {Serbyn},
  \citenamefont {Papi\ifmmode~\acute{c}\else \'{c}\fi{}},\ and\ \citenamefont
  {Abanin}}]{Serbyn2013}%
  \BibitemOpen
  \bibfield  {author} {\bibinfo {author} {\bibfnamefont {Maksym}\ \bibnamefont
  {Serbyn}}, \bibinfo {author} {\bibfnamefont {Z.}~\bibnamefont
  {Papi\ifmmode~\acute{c}\else \'{c}\fi{}}}, \ and\ \bibinfo {author}
  {\bibfnamefont {Dmitry~A.}\ \bibnamefont {Abanin}},\ }\bibfield  {title}
  {\enquote {\bibinfo {title} {Local conservation laws and the structure of the
  many-body localized states},}\ }\href {\doibase
  10.1103/PhysRevLett.111.127201} {\bibfield  {journal} {\bibinfo  {journal}
  {Phys. Rev. Lett.}\ }\textbf {\bibinfo {volume} {111}},\ \bibinfo {pages}
  {127201} (\bibinfo {year} {2013})}\BibitemShut {NoStop}%
\bibitem [{\citenamefont {Huse}\ \emph {et~al.}(2014)\citenamefont {Huse},
  \citenamefont {Nandkishore},\ and\ \citenamefont {Oganesyan}}]{Huse2014}%
  \BibitemOpen
  \bibfield  {author} {\bibinfo {author} {\bibfnamefont {David~A.}\
  \bibnamefont {Huse}}, \bibinfo {author} {\bibfnamefont {Rahul}\ \bibnamefont
  {Nandkishore}}, \ and\ \bibinfo {author} {\bibfnamefont {Vadim}\ \bibnamefont
  {Oganesyan}},\ }\bibfield  {title} {\enquote {\bibinfo {title} {Phenomenology
  of fully many-body-localized systems},}\ }\href {\doibase
  10.1103/PhysRevB.90.174202} {\bibfield  {journal} {\bibinfo  {journal} {Phys.
  Rev. B}\ }\textbf {\bibinfo {volume} {90}},\ \bibinfo {pages} {174202}
  (\bibinfo {year} {2014})}\BibitemShut {NoStop}%
\bibitem [{\citenamefont {Ponte}\ \emph
  {et~al.}(2015{\natexlab{a}})\citenamefont {Ponte}, \citenamefont
  {Papi\ifmmode~\acute{c}\else \'{c}\fi{}}, \citenamefont {Huveneers},\ and\
  \citenamefont {Abanin}}]{Ponte2015}%
  \BibitemOpen
  \bibfield  {author} {\bibinfo {author} {\bibfnamefont {Pedro}\ \bibnamefont
  {Ponte}}, \bibinfo {author} {\bibfnamefont {Z.}~\bibnamefont
  {Papi\ifmmode~\acute{c}\else \'{c}\fi{}}}, \bibinfo {author} {\bibfnamefont
  {Fran\ifmmode \mbox{\c{c}}\else~\c{c}\fi{}ois}\ \bibnamefont {Huveneers}}, \
  and\ \bibinfo {author} {\bibfnamefont {Dmitry~A.}\ \bibnamefont {Abanin}},\
  }\bibfield  {title} {\enquote {\bibinfo {title} {Many-body localization in
  periodically driven systems},}\ }\href {\doibase
  10.1103/PhysRevLett.114.140401} {\bibfield  {journal} {\bibinfo  {journal}
  {Phys. Rev. Lett.}\ }\textbf {\bibinfo {volume} {114}},\ \bibinfo {pages}
  {140401} (\bibinfo {year} {2015}{\natexlab{a}})}\BibitemShut {NoStop}%
\bibitem [{\citenamefont {Ponte}\ \emph
  {et~al.}(2015{\natexlab{b}})\citenamefont {Ponte}, \citenamefont {Chandran},
  \citenamefont {Papic},\ and\ \citenamefont {Abanin}}]{Ponte2015a}%
  \BibitemOpen
  \bibfield  {author} {\bibinfo {author} {\bibfnamefont {Pedro}\ \bibnamefont
  {Ponte}}, \bibinfo {author} {\bibfnamefont {Anushya}\ \bibnamefont
  {Chandran}}, \bibinfo {author} {\bibfnamefont {Z.}~\bibnamefont {Papic}}, \
  and\ \bibinfo {author} {\bibfnamefont {Dmitry~A.}\ \bibnamefont {Abanin}},\
  }\bibfield  {title} {\enquote {\bibinfo {title} {Periodically driven ergodic
  and many-body localized quantum systems},}\ }\href {\doibase
  10.1016/j.aop.2014.11.008} {\bibfield  {journal} {\bibinfo  {journal} {Annals
  of Physics}\ }\textbf {\bibinfo {volume} {353}},\ \bibinfo {pages} {196--204}
  (\bibinfo {year} {2015}{\natexlab{b}})}\BibitemShut {NoStop}%
\bibitem [{\citenamefont {Fishman}\ \emph {et~al.}(1982)\citenamefont
  {Fishman}, \citenamefont {Grempel},\ and\ \citenamefont
  {Prange}}]{Fishman1982}%
  \BibitemOpen
  \bibfield  {author} {\bibinfo {author} {\bibfnamefont {Shmuel}\ \bibnamefont
  {Fishman}}, \bibinfo {author} {\bibfnamefont {D.~R.}\ \bibnamefont
  {Grempel}}, \ and\ \bibinfo {author} {\bibfnamefont {R.~E.}\ \bibnamefont
  {Prange}},\ }\bibfield  {title} {\enquote {\bibinfo {title} {Chaos, quantum
  recurrences, and anderson localization},}\ }\href {\doibase
  10.1103/PhysRevLett.49.509} {\bibfield  {journal} {\bibinfo  {journal} {Phys.
  Rev. Lett.}\ }\textbf {\bibinfo {volume} {49}},\ \bibinfo {pages} {509--512}
  (\bibinfo {year} {1982})}\BibitemShut {NoStop}%
\bibitem [{\citenamefont {Chab\'e}\ \emph {et~al.}(2008)\citenamefont
  {Chab\'e}, \citenamefont {Lemari\'e}, \citenamefont {Gr\'emaud},
  \citenamefont {Delande}, \citenamefont {Szriftgiser},\ and\ \citenamefont
  {Garreau}}]{Chabe2008}%
  \BibitemOpen
  \bibfield  {author} {\bibinfo {author} {\bibfnamefont {Julien}\ \bibnamefont
  {Chab\'e}}, \bibinfo {author} {\bibfnamefont {Gabriel}\ \bibnamefont
  {Lemari\'e}}, \bibinfo {author} {\bibfnamefont {Benoit}\ \bibnamefont
  {Gr\'emaud}}, \bibinfo {author} {\bibfnamefont {Dominique}\ \bibnamefont
  {Delande}}, \bibinfo {author} {\bibfnamefont {Pascal}\ \bibnamefont
  {Szriftgiser}}, \ and\ \bibinfo {author} {\bibfnamefont {Jean~Claude}\
  \bibnamefont {Garreau}},\ }\bibfield  {title} {\enquote {\bibinfo {title}
  {Experimental observation of the anderson metal-insulator transition with
  atomic matter waves},}\ }\href {\doibase 10.1103/PhysRevLett.101.255702}
  {\bibfield  {journal} {\bibinfo  {journal} {Phys. Rev. Lett.}\ }\textbf
  {\bibinfo {volume} {101}},\ \bibinfo {pages} {255702} (\bibinfo {year}
  {2008})}\BibitemShut {NoStop}%
\bibitem [{\citenamefont {Lemari\'e}\ \emph {et~al.}(2010)\citenamefont
  {Lemari\'e}, \citenamefont {Lignier}, \citenamefont {Delande}, \citenamefont
  {Szriftgiser},\ and\ \citenamefont {Garreau}}]{Lemarie2010}%
  \BibitemOpen
  \bibfield  {author} {\bibinfo {author} {\bibfnamefont {Gabriel}\ \bibnamefont
  {Lemari\'e}}, \bibinfo {author} {\bibfnamefont {Hans}\ \bibnamefont
  {Lignier}}, \bibinfo {author} {\bibfnamefont {Dominique}\ \bibnamefont
  {Delande}}, \bibinfo {author} {\bibfnamefont {Pascal}\ \bibnamefont
  {Szriftgiser}}, \ and\ \bibinfo {author} {\bibfnamefont {Jean~Claude}\
  \bibnamefont {Garreau}},\ }\bibfield  {title} {\enquote {\bibinfo {title}
  {Critical state of the anderson transition: Between a metal and an
  insulator},}\ }\href {\doibase 10.1103/PhysRevLett.105.090601} {\bibfield
  {journal} {\bibinfo  {journal} {Phys. Rev. Lett.}\ }\textbf {\bibinfo
  {volume} {105}},\ \bibinfo {pages} {090601} (\bibinfo {year}
  {2010})}\BibitemShut {NoStop}%
\bibitem [{\citenamefont {Lopez}\ \emph {et~al.}(2012)\citenamefont {Lopez},
  \citenamefont {Cl\'ement}, \citenamefont {Szriftgiser}, \citenamefont
  {Garreau},\ and\ \citenamefont {Delande}}]{Lopez2012}%
  \BibitemOpen
  \bibfield  {author} {\bibinfo {author} {\bibfnamefont {Matthias}\
  \bibnamefont {Lopez}}, \bibinfo {author} {\bibfnamefont {Jean-Fran\ifmmode
  \mbox{\c{c}}\else~\c{c}\fi{}ois}\ \bibnamefont {Cl\'ement}}, \bibinfo
  {author} {\bibfnamefont {Pascal}\ \bibnamefont {Szriftgiser}}, \bibinfo
  {author} {\bibfnamefont {Jean~Claude}\ \bibnamefont {Garreau}}, \ and\
  \bibinfo {author} {\bibfnamefont {Dominique}\ \bibnamefont {Delande}},\
  }\bibfield  {title} {\enquote {\bibinfo {title} {Experimental test of
  universality of the anderson transition},}\ }\href {\doibase
  10.1103/PhysRevLett.108.095701} {\bibfield  {journal} {\bibinfo  {journal}
  {Phys. Rev. Lett.}\ }\textbf {\bibinfo {volume} {108}},\ \bibinfo {pages}
  {095701} (\bibinfo {year} {2012})}\BibitemShut {NoStop}%
\bibitem [{\citenamefont {Manai}\ \emph {et~al.}(2015)\citenamefont {Manai},
  \citenamefont {Cl\'ement}, \citenamefont {Chicireanu}, \citenamefont
  {Hainaut}, \citenamefont {Garreau}, \citenamefont {Szriftgiser},\ and\
  \citenamefont {Delande}}]{Manai2015}%
  \BibitemOpen
  \bibfield  {author} {\bibinfo {author} {\bibfnamefont {Isam}\ \bibnamefont
  {Manai}}, \bibinfo {author} {\bibfnamefont {Jean-Fran{\c{c}}ois}\
  \bibnamefont {Cl\'ement}}, \bibinfo {author} {\bibfnamefont {Radu}\
  \bibnamefont {Chicireanu}}, \bibinfo {author} {\bibfnamefont {Cl\'ement}\
  \bibnamefont {Hainaut}}, \bibinfo {author} {\bibfnamefont {Jean~Claude}\
  \bibnamefont {Garreau}}, \bibinfo {author} {\bibfnamefont {Pascal}\
  \bibnamefont {Szriftgiser}}, \ and\ \bibinfo {author} {\bibfnamefont
  {Dominique}\ \bibnamefont {Delande}},\ }\bibfield  {title} {\enquote
  {\bibinfo {title} {Experimental observation of two-dimensional anderson
  localization with the atomic kicked rotor},}\ }\href {\doibase
  10.1103/PhysRevLett.115.240603} {\bibfield  {journal} {\bibinfo  {journal}
  {Phys. Rev. Lett.}\ }\textbf {\bibinfo {volume} {115}},\ \bibinfo {pages}
  {240603} (\bibinfo {year} {2015})}\BibitemShut {NoStop}%
\bibitem [{\citenamefont {Hainaut}\ \emph
  {et~al.}(2018{\natexlab{a}})\citenamefont {Hainaut}, \citenamefont {Manai},
  \citenamefont {Cl{\'{e}}ment}, \citenamefont {Garreau}, \citenamefont
  {Szriftgiser}, \citenamefont {Lemari{\'{e}}}, \citenamefont {Cherroret},
  \citenamefont {Delande},\ and\ \citenamefont {Chicireanu}}]{Hainaut2018b}%
  \BibitemOpen
  \bibfield  {author} {\bibinfo {author} {\bibfnamefont {Cl{\'{e}}ment}\
  \bibnamefont {Hainaut}}, \bibinfo {author} {\bibfnamefont {Isam}\
  \bibnamefont {Manai}}, \bibinfo {author} {\bibfnamefont
  {Jean-Fran{\c{c}}ois}\ \bibnamefont {Cl{\'{e}}ment}}, \bibinfo {author}
  {\bibfnamefont {Jean~Claude}\ \bibnamefont {Garreau}}, \bibinfo {author}
  {\bibfnamefont {Pascal}\ \bibnamefont {Szriftgiser}}, \bibinfo {author}
  {\bibfnamefont {Gabriel}\ \bibnamefont {Lemari{\'{e}}}}, \bibinfo {author}
  {\bibfnamefont {Nicolas}\ \bibnamefont {Cherroret}}, \bibinfo {author}
  {\bibfnamefont {Dominique}\ \bibnamefont {Delande}}, \ and\ \bibinfo {author}
  {\bibfnamefont {Radu}\ \bibnamefont {Chicireanu}},\ }\bibfield  {title}
  {\enquote {\bibinfo {title} {Controlling symmetry and localization with an
  artificial gauge field in a disordered quantum system},}\ }\href {\doibase
  10.1038/s41467-018-03481-9} {\bibfield  {journal} {\bibinfo  {journal}
  {Nature Communications}\ }\textbf {\bibinfo {volume} {9}},\ \bibinfo {pages}
  {1382} (\bibinfo {year} {2018}{\natexlab{a}})}\BibitemShut {NoStop}%
\bibitem [{\citenamefont {Hainaut}\ \emph
  {et~al.}(2018{\natexlab{b}})\citenamefont {Hainaut}, \citenamefont {Fang},
  \citenamefont {Ran\ifmmode~\mbox{\c{c}}\else \c{c}\fi{}on}, \citenamefont
  {Cl\'ement}, \citenamefont {Szriftgiser}, \citenamefont {Garreau},
  \citenamefont {Tian},\ and\ \citenamefont {Chicireanu}}]{Hainaut2018a}%
  \BibitemOpen
  \bibfield  {author} {\bibinfo {author} {\bibfnamefont {Cl\'ement}\
  \bibnamefont {Hainaut}}, \bibinfo {author} {\bibfnamefont {Ping}\
  \bibnamefont {Fang}}, \bibinfo {author} {\bibfnamefont {Adam}\ \bibnamefont
  {Ran\ifmmode~\mbox{\c{c}}\else \c{c}\fi{}on}}, \bibinfo {author}
  {\bibfnamefont {Jean-Francois}\ \bibnamefont {Cl\'ement}}, \bibinfo {author}
  {\bibfnamefont {Pascal}\ \bibnamefont {Szriftgiser}}, \bibinfo {author}
  {\bibfnamefont {Jean-Claude}\ \bibnamefont {Garreau}}, \bibinfo {author}
  {\bibfnamefont {Chushun}\ \bibnamefont {Tian}}, \ and\ \bibinfo {author}
  {\bibfnamefont {Radu}\ \bibnamefont {Chicireanu}},\ }\bibfield  {title}
  {\enquote {\bibinfo {title} {Experimental observation of a time-driven phase
  transition in quantum chaos},}\ }\href {\doibase
  10.1103/PhysRevLett.121.134101} {\bibfield  {journal} {\bibinfo  {journal}
  {Phys. Rev. Lett.}\ }\textbf {\bibinfo {volume} {121}},\ \bibinfo {pages}
  {134101} (\bibinfo {year} {2018}{\natexlab{b}})}\BibitemShut {NoStop}%
\bibitem [{\citenamefont {Adachi}\ \emph {et~al.}(1988)\citenamefont {Adachi},
  \citenamefont {Toda},\ and\ \citenamefont {Ikeda}}]{Adachi1988}%
  \BibitemOpen
  \bibfield  {author} {\bibinfo {author} {\bibfnamefont {S.}~\bibnamefont
  {Adachi}}, \bibinfo {author} {\bibfnamefont {M.}~\bibnamefont {Toda}}, \ and\
  \bibinfo {author} {\bibfnamefont {K.}~\bibnamefont {Ikeda}},\ }\bibfield
  {title} {\enquote {\bibinfo {title} {Quantum-classical correspondence in
  many-dimensional quantum chaos},}\ }\href {\doibase
  10.1103/PhysRevLett.61.659} {\bibfield  {journal} {\bibinfo  {journal} {Phys.
  Rev. Lett.}\ }\textbf {\bibinfo {volume} {61}},\ \bibinfo {pages} {659--661}
  (\bibinfo {year} {1988})}\BibitemShut {NoStop}%
\bibitem [{\citenamefont {Wen-Lei}\ and\ \citenamefont
  {Quan-Lin}(2009)}]{Lei2009}%
  \BibitemOpen
  \bibfield  {author} {\bibinfo {author} {\bibfnamefont {Zhao}\ \bibnamefont
  {Wen-Lei}}\ and\ \bibinfo {author} {\bibfnamefont {Jie}\ \bibnamefont
  {Quan-Lin}},\ }\bibfield  {title} {\enquote {\bibinfo {title} {Quantum to
  classical transition in a system of two coupled kicked rotors},}\ }\href
  {\doibase 10.1088/0253-6102/51/3/17} {\bibfield  {journal} {\bibinfo
  {journal} {Communications in Theoretical Physics}\ }\textbf {\bibinfo
  {volume} {51}},\ \bibinfo {pages} {465--469} (\bibinfo {year}
  {2009})}\BibitemShut {NoStop}%
\bibitem [{\citenamefont {Keser}\ \emph {et~al.}(2016)\citenamefont {Keser},
  \citenamefont {Ganeshan}, \citenamefont {Refael},\ and\ \citenamefont
  {Galitski}}]{Keser2016}%
  \BibitemOpen
  \bibfield  {author} {\bibinfo {author} {\bibfnamefont {Aydin~Cem}\
  \bibnamefont {Keser}}, \bibinfo {author} {\bibfnamefont {Sriram}\
  \bibnamefont {Ganeshan}}, \bibinfo {author} {\bibfnamefont {Gil}\
  \bibnamefont {Refael}}, \ and\ \bibinfo {author} {\bibfnamefont {Victor}\
  \bibnamefont {Galitski}},\ }\bibfield  {title} {\enquote {\bibinfo {title}
  {Dynamical many-body localization in an integrable model},}\ }\href {\doibase
  10.1103/PhysRevB.94.085120} {\bibfield  {journal} {\bibinfo  {journal} {Phys.
  Rev. B}\ }\textbf {\bibinfo {volume} {94}},\ \bibinfo {pages} {085120}
  (\bibinfo {year} {2016})}\BibitemShut {NoStop}%
\bibitem [{\citenamefont {Rozenbaum}\ and\ \citenamefont
  {Galitski}(2017)}]{Rozenbaum2017}%
  \BibitemOpen
  \bibfield  {author} {\bibinfo {author} {\bibfnamefont {Efim~B.}\ \bibnamefont
  {Rozenbaum}}\ and\ \bibinfo {author} {\bibfnamefont {Victor}\ \bibnamefont
  {Galitski}},\ }\bibfield  {title} {\enquote {\bibinfo {title} {Dynamical
  localization of coupled relativistic kicked rotors},}\ }\href {\doibase
  10.1103/PhysRevB.95.064303} {\bibfield  {journal} {\bibinfo  {journal} {Phys.
  Rev. B}\ }\textbf {\bibinfo {volume} {95}},\ \bibinfo {pages} {064303}
  (\bibinfo {year} {2017})}\BibitemShut {NoStop}%
\bibitem [{\citenamefont {Notarnicola}\ \emph {et~al.}(2018)\citenamefont
  {Notarnicola}, \citenamefont {Iemini}, \citenamefont {Rossini}, \citenamefont
  {Fazio}, \citenamefont {Silva},\ and\ \citenamefont
  {Russomanno}}]{Notarnicola2018}%
  \BibitemOpen
  \bibfield  {author} {\bibinfo {author} {\bibfnamefont {Simone}\ \bibnamefont
  {Notarnicola}}, \bibinfo {author} {\bibfnamefont {Fernando}\ \bibnamefont
  {Iemini}}, \bibinfo {author} {\bibfnamefont {Davide}\ \bibnamefont
  {Rossini}}, \bibinfo {author} {\bibfnamefont {Rosario}\ \bibnamefont
  {Fazio}}, \bibinfo {author} {\bibfnamefont {Alessandro}\ \bibnamefont
  {Silva}}, \ and\ \bibinfo {author} {\bibfnamefont {Angelo}\ \bibnamefont
  {Russomanno}},\ }\bibfield  {title} {\enquote {\bibinfo {title} {From
  localization to anomalous diffusion in the dynamics of coupled kicked
  rotors},}\ }\href {\doibase 10.1103/PhysRevE.97.022202} {\bibfield  {journal}
  {\bibinfo  {journal} {Phys. Rev. E}\ }\textbf {\bibinfo {volume} {97}},\
  \bibinfo {pages} {022202} (\bibinfo {year} {2018})}\BibitemShut {NoStop}%
\bibitem [{\citenamefont {Notarnicola}\ \emph {et~al.}(2020)\citenamefont
  {Notarnicola}, \citenamefont {Silva}, \citenamefont {Fazio},\ and\
  \citenamefont {Russomanno}}]{Notarnicola2020}%
  \BibitemOpen
  \bibfield  {author} {\bibinfo {author} {\bibfnamefont {Simone}\ \bibnamefont
  {Notarnicola}}, \bibinfo {author} {\bibfnamefont {Alessandro}\ \bibnamefont
  {Silva}}, \bibinfo {author} {\bibfnamefont {Rosario}\ \bibnamefont {Fazio}},
  \ and\ \bibinfo {author} {\bibfnamefont {Angelo}\ \bibnamefont
  {Russomanno}},\ }\bibfield  {title} {\enquote {\bibinfo {title} {Slow heating
  in a quantum coupled kicked rotors system},}\ }\href {\doibase
  10.1088/1742-5468/ab6de4} {\bibfield  {journal} {\bibinfo  {journal} {Journal
  of Statistical Mechanics: Theory and Experiment}\ }\textbf {\bibinfo {volume}
  {2020}},\ \bibinfo {pages} {024008} (\bibinfo {year} {2020})}\BibitemShut
  {NoStop}%
\bibitem [{\citenamefont {Bloch}\ \emph {et~al.}(2008)\citenamefont {Bloch},
  \citenamefont {Dalibard},\ and\ \citenamefont {Zwerger}}]{Bloch2008}%
  \BibitemOpen
  \bibfield  {author} {\bibinfo {author} {\bibfnamefont {Immanuel}\
  \bibnamefont {Bloch}}, \bibinfo {author} {\bibfnamefont {Jean}\ \bibnamefont
  {Dalibard}}, \ and\ \bibinfo {author} {\bibfnamefont {Wilhelm}\ \bibnamefont
  {Zwerger}},\ }\bibfield  {title} {\enquote {\bibinfo {title} {Many-body
  physics with ultracold gases},}\ }\href {\doibase 10.1103/RevModPhys.80.885}
  {\bibfield  {journal} {\bibinfo  {journal} {Rev. Mod. Phys.}\ }\textbf
  {\bibinfo {volume} {80}},\ \bibinfo {pages} {885--964} (\bibinfo {year}
  {2008})}\BibitemShut {NoStop}%
\bibitem [{\citenamefont {Shepelyansky}(1993)}]{Shepelyansky1993}%
  \BibitemOpen
  \bibfield  {author} {\bibinfo {author} {\bibfnamefont {D.~L.}\ \bibnamefont
  {Shepelyansky}},\ }\bibfield  {title} {\enquote {\bibinfo {title}
  {Delocalization of quantum chaos by weak nonlinearity},}\ }\href {\doibase
  10.1103/PhysRevLett.70.1787} {\bibfield  {journal} {\bibinfo  {journal}
  {Phys. Rev. Lett.}\ }\textbf {\bibinfo {volume} {70}},\ \bibinfo {pages}
  {1787--1790} (\bibinfo {year} {1993})}\BibitemShut {NoStop}%
\bibitem [{\citenamefont {Pikovsky}\ and\ \citenamefont
  {Shepelyansky}(2008)}]{Pikovsky2008}%
  \BibitemOpen
  \bibfield  {author} {\bibinfo {author} {\bibfnamefont {A.~S.}\ \bibnamefont
  {Pikovsky}}\ and\ \bibinfo {author} {\bibfnamefont {D.~L.}\ \bibnamefont
  {Shepelyansky}},\ }\bibfield  {title} {\enquote {\bibinfo {title}
  {Destruction of anderson localization by a weak nonlinearity},}\ }\href
  {\doibase 10.1103/PhysRevLett.100.094101} {\bibfield  {journal} {\bibinfo
  {journal} {Phys. Rev. Lett.}\ }\textbf {\bibinfo {volume} {100}},\ \bibinfo
  {pages} {094101} (\bibinfo {year} {2008})}\BibitemShut {NoStop}%
\bibitem [{\citenamefont {Flach}\ \emph {et~al.}(2009)\citenamefont {Flach},
  \citenamefont {Krimer},\ and\ \citenamefont {Skokos}}]{Flach2009}%
  \BibitemOpen
  \bibfield  {author} {\bibinfo {author} {\bibfnamefont {S.}~\bibnamefont
  {Flach}}, \bibinfo {author} {\bibfnamefont {D.~O.}\ \bibnamefont {Krimer}}, \
  and\ \bibinfo {author} {\bibfnamefont {Ch.}\ \bibnamefont {Skokos}},\
  }\bibfield  {title} {\enquote {\bibinfo {title} {Universal spreading of wave
  packets in disordered nonlinear systems},}\ }\href {\doibase
  10.1103/PhysRevLett.102.024101} {\bibfield  {journal} {\bibinfo  {journal}
  {Phys. Rev. Lett.}\ }\textbf {\bibinfo {volume} {102}},\ \bibinfo {pages}
  {024101} (\bibinfo {year} {2009})}\BibitemShut {NoStop}%
\bibitem [{\citenamefont {Gligoric{\'a}}\ \emph {et~al.}(2011)\citenamefont
  {Gligoric{\'a}}, \citenamefont {Bodyfelt},\ and\ \citenamefont
  {Flach}}]{Gligoric2011}%
  \BibitemOpen
  \bibfield  {author} {\bibinfo {author} {\bibfnamefont {G.}~\bibnamefont
  {Gligoric{\'a}}}, \bibinfo {author} {\bibfnamefont {J.~D.}\ \bibnamefont
  {Bodyfelt}}, \ and\ \bibinfo {author} {\bibfnamefont {S.}~\bibnamefont
  {Flach}},\ }\bibfield  {title} {\enquote {\bibinfo {title} {Interactions
  destroy dynamical localization with strong and weak chaos},}\ }\href
  {http://stacks.iop.org/0295-5075/96/i=3/a=30004} {\bibfield  {journal}
  {\bibinfo  {journal} {EPL (Europhysics Letters)}\ }\textbf {\bibinfo {volume}
  {96}},\ \bibinfo {pages} {30004} (\bibinfo {year} {2011})}\BibitemShut
  {NoStop}%
\bibitem [{\citenamefont {Cherroret}\ \emph {et~al.}(2014)\citenamefont
  {Cherroret}, \citenamefont {Vermersch}, \citenamefont {Garreau},\ and\
  \citenamefont {Delande}}]{Cherroret2014}%
  \BibitemOpen
  \bibfield  {author} {\bibinfo {author} {\bibfnamefont {Nicolas}\ \bibnamefont
  {Cherroret}}, \bibinfo {author} {\bibfnamefont {Benoit}\ \bibnamefont
  {Vermersch}}, \bibinfo {author} {\bibfnamefont {Jean~Claude}\ \bibnamefont
  {Garreau}}, \ and\ \bibinfo {author} {\bibfnamefont {Dominique}\ \bibnamefont
  {Delande}},\ }\bibfield  {title} {\enquote {\bibinfo {title} {How nonlinear
  interactions challenge the three-dimensional anderson transition},}\ }\href
  {\doibase 10.1103/PhysRevLett.112.170603} {\bibfield  {journal} {\bibinfo
  {journal} {Phys. Rev. Lett.}\ }\textbf {\bibinfo {volume} {112}},\ \bibinfo
  {pages} {170603} (\bibinfo {year} {2014})}\BibitemShut {NoStop}%
\bibitem [{\citenamefont {Lellouch}\ \emph {et~al.}(2020)\citenamefont
  {Lellouch}, \citenamefont {Ran\ifmmode~\mbox{\c{c}}\else \c{c}\fi{}on},
  \citenamefont {De~Bi\`evre}, \citenamefont {Delande},\ and\ \citenamefont
  {Garreau}}]{Lellouch2020}%
  \BibitemOpen
  \bibfield  {author} {\bibinfo {author} {\bibfnamefont {Samuel}\ \bibnamefont
  {Lellouch}}, \bibinfo {author} {\bibfnamefont {Adam}\ \bibnamefont
  {Ran\ifmmode~\mbox{\c{c}}\else \c{c}\fi{}on}}, \bibinfo {author}
  {\bibfnamefont {Stephan}\ \bibnamefont {De~Bi\`evre}}, \bibinfo {author}
  {\bibfnamefont {Dominique}\ \bibnamefont {Delande}}, \ and\ \bibinfo {author}
  {\bibfnamefont {Jean~Claude}\ \bibnamefont {Garreau}},\ }\bibfield  {title}
  {\enquote {\bibinfo {title} {Dynamics of the mean-field-interacting quantum
  kicked rotor},}\ }\href {\doibase 10.1103/PhysRevA.101.043624} {\bibfield
  {journal} {\bibinfo  {journal} {Phys. Rev. A}\ }\textbf {\bibinfo {volume}
  {101}},\ \bibinfo {pages} {043624} (\bibinfo {year} {2020})}\BibitemShut
  {NoStop}%
\bibitem [{\citenamefont {Cazalilla}\ \emph {et~al.}(2011)\citenamefont
  {Cazalilla}, \citenamefont {Citro}, \citenamefont {Giamarchi}, \citenamefont
  {Orignac},\ and\ \citenamefont {Rigol}}]{Cazalilla2011}%
  \BibitemOpen
  \bibfield  {author} {\bibinfo {author} {\bibfnamefont {M.~A.}\ \bibnamefont
  {Cazalilla}}, \bibinfo {author} {\bibfnamefont {R.}~\bibnamefont {Citro}},
  \bibinfo {author} {\bibfnamefont {T.}~\bibnamefont {Giamarchi}}, \bibinfo
  {author} {\bibfnamefont {E.}~\bibnamefont {Orignac}}, \ and\ \bibinfo
  {author} {\bibfnamefont {M.}~\bibnamefont {Rigol}},\ }\bibfield  {title}
  {\enquote {\bibinfo {title} {One dimensional bosons: From condensed matter
  systems to ultracold gases},}\ }\href {\doibase 10.1103/RevModPhys.83.1405}
  {\bibfield  {journal} {\bibinfo  {journal} {Reviews of Modern Physics}\
  }\textbf {\bibinfo {volume} {83}},\ \bibinfo {pages} {1405--1466} (\bibinfo
  {year} {2011})}\BibitemShut {NoStop}%
\bibitem [{\citenamefont {Qin}\ \emph {et~al.}(2017)\citenamefont {Qin},
  \citenamefont {Andreanov}, \citenamefont {Park},\ and\ \citenamefont
  {Flach}}]{Qin2017}%
  \BibitemOpen
  \bibfield  {author} {\bibinfo {author} {\bibfnamefont {Pinquan}\ \bibnamefont
  {Qin}}, \bibinfo {author} {\bibfnamefont {Alexei}\ \bibnamefont {Andreanov}},
  \bibinfo {author} {\bibfnamefont {Hee~Chul}\ \bibnamefont {Park}}, \ and\
  \bibinfo {author} {\bibfnamefont {Sergej}\ \bibnamefont {Flach}},\ }\bibfield
   {title} {\enquote {\bibinfo {title} {Interacting ultracold atomic kicked
  rotors: loss of dynamical localization},}\ }\href {\doibase
  10.1038/srep41139} {\bibfield  {journal} {\bibinfo  {journal} {Scientific
  Reports}\ }\textbf {\bibinfo {volume} {7}},\ \bibinfo {pages} {41139}
  (\bibinfo {year} {2017})}\BibitemShut {NoStop}%
\bibitem [{\citenamefont {Chicireanu}\ and\ \citenamefont
  {Ran\ifmmode~\mbox{\c{c}}\else \c{c}\fi{}on}(2021)}]{Chicireanu2021}%
  \BibitemOpen
  \bibfield  {author} {\bibinfo {author} {\bibfnamefont {Radu}\ \bibnamefont
  {Chicireanu}}\ and\ \bibinfo {author} {\bibfnamefont {Adam}\ \bibnamefont
  {Ran\ifmmode~\mbox{\c{c}}\else \c{c}\fi{}on}},\ }\bibfield  {title} {\enquote
  {\bibinfo {title} {Dynamical localization of interacting bosons in the
  few-body limit},}\ }\href {\doibase 10.1103/PhysRevA.103.043314} {\bibfield
  {journal} {\bibinfo  {journal} {Phys. Rev. A}\ }\textbf {\bibinfo {volume}
  {103}},\ \bibinfo {pages} {043314} (\bibinfo {year} {2021})}\BibitemShut
  {NoStop}%
\bibitem [{\citenamefont {Rylands}\ \emph {et~al.}(2020)\citenamefont
  {Rylands}, \citenamefont {Rozenbaum}, \citenamefont {Galitski},\ and\
  \citenamefont {Konik}}]{Rylands2020}%
  \BibitemOpen
  \bibfield  {author} {\bibinfo {author} {\bibfnamefont {Colin}\ \bibnamefont
  {Rylands}}, \bibinfo {author} {\bibfnamefont {Efim~B.}\ \bibnamefont
  {Rozenbaum}}, \bibinfo {author} {\bibfnamefont {Victor}\ \bibnamefont
  {Galitski}}, \ and\ \bibinfo {author} {\bibfnamefont {Robert}\ \bibnamefont
  {Konik}},\ }\bibfield  {title} {\enquote {\bibinfo {title} {Many-body
  dynamical localization in a kicked lieb-liniger gas},}\ }\href {\doibase
  10.1103/PhysRevLett.124.155302} {\bibfield  {journal} {\bibinfo  {journal}
  {Phys. Rev. Lett.}\ }\textbf {\bibinfo {volume} {124}},\ \bibinfo {pages}
  {155302} (\bibinfo {year} {2020})}\BibitemShut {NoStop}%
\bibitem [{\citenamefont {Lazarides}\ \emph {et~al.}(2014)\citenamefont
  {Lazarides}, \citenamefont {Das},\ and\ \citenamefont
  {Moessner}}]{Lazarides2014}%
  \BibitemOpen
  \bibfield  {author} {\bibinfo {author} {\bibfnamefont {Achilleas}\
  \bibnamefont {Lazarides}}, \bibinfo {author} {\bibfnamefont {Arnab}\
  \bibnamefont {Das}}, \ and\ \bibinfo {author} {\bibfnamefont {Roderich}\
  \bibnamefont {Moessner}},\ }\bibfield  {title} {\enquote {\bibinfo {title}
  {Periodic thermodynamics of isolated quantum systems},}\ }\href {\doibase
  10.1103/PhysRevLett.112.150401} {\bibfield  {journal} {\bibinfo  {journal}
  {Phys. Rev. Lett.}\ }\textbf {\bibinfo {volume} {112}},\ \bibinfo {pages}
  {150401} (\bibinfo {year} {2014})}\BibitemShut {NoStop}%
\bibitem [{\citenamefont {Vidmar}\ and\ \citenamefont
  {Rigol}(2016)}]{Vidmar2016}%
  \BibitemOpen
  \bibfield  {author} {\bibinfo {author} {\bibfnamefont {Lev}\ \bibnamefont
  {Vidmar}}\ and\ \bibinfo {author} {\bibfnamefont {Marcos}\ \bibnamefont
  {Rigol}},\ }\bibfield  {title} {\enquote {\bibinfo {title} {Generalized gibbs
  ensemble in integrable lattice models},}\ }\href {\doibase
  10.1088/1742-5468/2016/06/064007} {\bibfield  {journal} {\bibinfo  {journal}
  {Journal of Statistical Mechanics: Theory and Experiment}\ }\textbf {\bibinfo
  {volume} {2016}},\ \bibinfo {pages} {064007} (\bibinfo {year}
  {2016})}\BibitemShut {NoStop}%
\bibitem [{\citenamefont {Olshanii}\ and\ \citenamefont
  {Dunjko}(2003)}]{Olshanii2003}%
  \BibitemOpen
  \bibfield  {author} {\bibinfo {author} {\bibfnamefont {Maxim}\ \bibnamefont
  {Olshanii}}\ and\ \bibinfo {author} {\bibfnamefont {Vanja}\ \bibnamefont
  {Dunjko}},\ }\bibfield  {title} {\enquote {\bibinfo {title} {Short-distance
  correlation properties of the lieb-liniger system and momentum distributions
  of trapped one-dimensional atomic gases},}\ }\href {\doibase
  10.1103/PhysRevLett.91.090401} {\bibfield  {journal} {\bibinfo  {journal}
  {Physical Review Letters}\ }\textbf {\bibinfo {volume} {91}},\ \bibinfo
  {pages} {090401--} (\bibinfo {year} {2003})}\BibitemShut {NoStop}%
\bibitem [{\citenamefont {Tan}(2008)}]{Tan2008}%
  \BibitemOpen
  \bibfield  {author} {\bibinfo {author} {\bibfnamefont {Shina}\ \bibnamefont
  {Tan}},\ }\bibfield  {title} {\enquote {\bibinfo {title} {Large momentum part
  of a strongly correlated fermi gas},}\ }\href {\doibase
  http://dx.doi.org/10.1016/j.aop.2008.03.005} {\bibfield  {journal} {\bibinfo
  {journal} {Annals of Physics}\ }\textbf {\bibinfo {volume} {323}},\ \bibinfo
  {pages} {2971 -- 2986} (\bibinfo {year} {2008})}\BibitemShut {NoStop}%
\bibitem [{\citenamefont {Shepelyansky}(1986)}]{Shepelyansky1986}%
  \BibitemOpen
  \bibfield  {author} {\bibinfo {author} {\bibfnamefont {D.~L.}\ \bibnamefont
  {Shepelyansky}},\ }\bibfield  {title} {\enquote {\bibinfo {title}
  {Localization of quasienergy eigenfunctions in action space},}\ }\href
  {\doibase 10.1103/PhysRevLett.56.677} {\bibfield  {journal} {\bibinfo
  {journal} {Phys. Rev. Lett.}\ }\textbf {\bibinfo {volume} {56}},\ \bibinfo
  {pages} {677--680} (\bibinfo {year} {1986})}\BibitemShut {NoStop}%
\bibitem [{\citenamefont {Lemari\'e}\ \emph {et~al.}(2009)\citenamefont
  {Lemari\'e}, \citenamefont {Chab\'e}, \citenamefont {Szriftgiser},
  \citenamefont {Garreau}, \citenamefont {Gr\'emaud},\ and\ \citenamefont
  {Delande}}]{Lemarie2009}%
  \BibitemOpen
  \bibfield  {author} {\bibinfo {author} {\bibfnamefont {Gabriel}\ \bibnamefont
  {Lemari\'e}}, \bibinfo {author} {\bibfnamefont {Julien}\ \bibnamefont
  {Chab\'e}}, \bibinfo {author} {\bibfnamefont {Pascal}\ \bibnamefont
  {Szriftgiser}}, \bibinfo {author} {\bibfnamefont {Jean~Claude}\ \bibnamefont
  {Garreau}}, \bibinfo {author} {\bibfnamefont {Benoit}\ \bibnamefont
  {Gr\'emaud}}, \ and\ \bibinfo {author} {\bibfnamefont {Dominique}\
  \bibnamefont {Delande}},\ }\bibfield  {title} {\enquote {\bibinfo {title}
  {Observation of the anderson metal-insulator transition with atomic matter
  waves: Theory and experiment},}\ }\href {\doibase 10.1103/PhysRevA.80.043626}
  {\bibfield  {journal} {\bibinfo  {journal} {Phys. Rev. A}\ }\textbf {\bibinfo
  {volume} {80}},\ \bibinfo {pages} {043626} (\bibinfo {year}
  {2009})}\BibitemShut {NoStop}%
\bibitem [{Note1()}]{Note1}%
  \BibitemOpen
  \bibinfo {note} {To be precise, we define $p_{loc}$ of a single-particle
  state $ | \psi \rangle $ as $p_{loc}^2=\langle \psi |\hat p^2|\psi \rangle
  -\langle \psi |\hat p|\psi \rangle ^2$, which is proportional to the momentum
  scale on which the dynamically localized states decay
  exponentially.}\BibitemShut {Stop}%
\bibitem [{\citenamefont {Girardeau}(1960)}]{Girardeau1960}%
  \BibitemOpen
  \bibfield  {author} {\bibinfo {author} {\bibfnamefont {M.}~\bibnamefont
  {Girardeau}},\ }\bibfield  {title} {\enquote {\bibinfo {title} {Relationship
  between systems of impenetrable bosons and fermions in one dimension},}\
  }\href {\doibase 10.1063/1.1703687} {\bibfield  {journal} {\bibinfo
  {journal} {Journal of Mathematical Physics}\ }\textbf {\bibinfo {volume}
  {1}},\ \bibinfo {pages} {516--523} (\bibinfo {year} {1960})}\BibitemShut
  {NoStop}%
\bibitem [{\citenamefont {Lenard}(1964)}]{Lenard1964}%
  \BibitemOpen
  \bibfield  {author} {\bibinfo {author} {\bibfnamefont {A.}~\bibnamefont
  {Lenard}},\ }\bibfield  {title} {\enquote {\bibinfo {title} {Momentum
  distribution in the ground state of the one‐dimensional system of
  impenetrable bosons},}\ }\href {\doibase 10.1063/1.1704196} {\bibfield
  {journal} {\bibinfo  {journal} {Journal of Mathematical Physics}\ }\textbf
  {\bibinfo {volume} {5}},\ \bibinfo {pages} {930--943} (\bibinfo {year}
  {1964})}\BibitemShut {NoStop}%
\bibitem [{\citenamefont {Buljan}\ \emph {et~al.}(2008)\citenamefont {Buljan},
  \citenamefont {Pezer},\ and\ \citenamefont {Gasenzer}}]{Buljan2008}%
  \BibitemOpen
  \bibfield  {author} {\bibinfo {author} {\bibfnamefont {H.}~\bibnamefont
  {Buljan}}, \bibinfo {author} {\bibfnamefont {R.}~\bibnamefont {Pezer}}, \
  and\ \bibinfo {author} {\bibfnamefont {T.}~\bibnamefont {Gasenzer}},\
  }\bibfield  {title} {\enquote {\bibinfo {title} {Fermi-bose transformation
  for the time-dependent lieb-liniger gas},}\ }\href {\doibase
  10.1103/PhysRevLett.100.080406} {\bibfield  {journal} {\bibinfo  {journal}
  {Phys. Rev. Lett.}\ }\textbf {\bibinfo {volume} {100}},\ \bibinfo {pages}
  {080406} (\bibinfo {year} {2008})}\BibitemShut {NoStop}%
\bibitem [{\citenamefont {Juki{\'c}}\ \emph {et~al.}(2008)\citenamefont
  {Juki{\'c}}, \citenamefont {Pezer}, \citenamefont {Gasenzer},\ and\
  \citenamefont {Buljan}}]{Jukic2008}%
  \BibitemOpen
  \bibfield  {author} {\bibinfo {author} {\bibfnamefont {D.}~\bibnamefont
  {Juki{\'c}}}, \bibinfo {author} {\bibfnamefont {R.}~\bibnamefont {Pezer}},
  \bibinfo {author} {\bibfnamefont {T.}~\bibnamefont {Gasenzer}}, \ and\
  \bibinfo {author} {\bibfnamefont {H.}~\bibnamefont {Buljan}},\ }\bibfield
  {title} {\enquote {\bibinfo {title} {Free expansion of a lieb-liniger gas:
  Asymptotic form of the wave functions},}\ }\href {\doibase
  10.1103/PhysRevA.78.053602} {\bibfield  {journal} {\bibinfo  {journal}
  {Physical Review A}\ }\textbf {\bibinfo {volume} {78}},\ \bibinfo {pages}
  {053602--} (\bibinfo {year} {2008})}\BibitemShut {NoStop}%
\bibitem [{\citenamefont {Pezer}\ \emph {et~al.}(2009)\citenamefont {Pezer},
  \citenamefont {Gasenzer},\ and\ \citenamefont {Buljan}}]{Pezer2009}%
  \BibitemOpen
  \bibfield  {author} {\bibinfo {author} {\bibfnamefont {R.}~\bibnamefont
  {Pezer}}, \bibinfo {author} {\bibfnamefont {T.}~\bibnamefont {Gasenzer}}, \
  and\ \bibinfo {author} {\bibfnamefont {H.}~\bibnamefont {Buljan}},\
  }\bibfield  {title} {\enquote {\bibinfo {title} {Single-particle density
  matrix for a time-dependent strongly interacting one-dimensional bose gas},}\
  }\href {\doibase 10.1103/PhysRevA.80.053616} {\bibfield  {journal} {\bibinfo
  {journal} {Physical Review A}\ }\textbf {\bibinfo {volume} {80}},\ \bibinfo
  {pages} {053616--} (\bibinfo {year} {2009})}\BibitemShut {NoStop}%
\bibitem [{Note2()}]{Note2}%
  \BibitemOpen
  \bibinfo {note} {This is to be contrasted with the case of a Tonks gas loaded
  in a random potential, where localization in real space of the fermions is
  naturally preserved for the bosons \cite
  {Radic2010,Seiringer2016}.}\BibitemShut {Stop}%
\bibitem [{\citenamefont {Rigol}\ and\ \citenamefont
  {Muramatsu}(2005{\natexlab{a}})}]{Rigol2005a}%
  \BibitemOpen
  \bibfield  {author} {\bibinfo {author} {\bibfnamefont {Marcos}\ \bibnamefont
  {Rigol}}\ and\ \bibinfo {author} {\bibfnamefont {Alejandro}\ \bibnamefont
  {Muramatsu}},\ }\bibfield  {title} {\enquote {\bibinfo {title}
  {Fermionization in an expanding 1d gas of hard-core bosons},}\ }\href
  {\doibase 10.1103/PhysRevLett.94.240403} {\bibfield  {journal} {\bibinfo
  {journal} {Physical Review Letters}\ }\textbf {\bibinfo {volume} {94}},\
  \bibinfo {pages} {240403--} (\bibinfo {year}
  {2005}{\natexlab{a}})}\BibitemShut {NoStop}%
\bibitem [{\citenamefont {Rigol}\ and\ \citenamefont
  {Muramatsu}(2005{\natexlab{b}})}]{Rigol2005}%
  \BibitemOpen
  \bibfield  {author} {\bibinfo {author} {\bibfnamefont {Marcos}\ \bibnamefont
  {Rigol}}\ and\ \bibinfo {author} {\bibfnamefont {Alejandro}\ \bibnamefont
  {Muramatsu}},\ }\bibfield  {title} {\enquote {\bibinfo {title} {Ground-state
  properties of hard-core bosons confined on one-dimensional optical
  lattices},}\ }\href {\doibase 10.1103/PhysRevA.72.013604} {\bibfield
  {journal} {\bibinfo  {journal} {Physical Review A}\ }\textbf {\bibinfo
  {volume} {72}},\ \bibinfo {pages} {013604--} (\bibinfo {year}
  {2005}{\natexlab{b}})}\BibitemShut {NoStop}%
\bibitem [{\citenamefont {Vaidya}\ and\ \citenamefont
  {Tracy}(1979)}]{Vaidya1979}%
  \BibitemOpen
  \bibfield  {author} {\bibinfo {author} {\bibfnamefont {H.~G.}\ \bibnamefont
  {Vaidya}}\ and\ \bibinfo {author} {\bibfnamefont {C.~A.}\ \bibnamefont
  {Tracy}},\ }\bibfield  {title} {\enquote {\bibinfo {title} {One-particle
  reduced density matrix of impenetrable bosons in one dimension at zero
  temperature},}\ }\href {\doibase 10.1103/PhysRevLett.42.3} {\bibfield
  {journal} {\bibinfo  {journal} {Phys. Rev. Lett.}\ }\textbf {\bibinfo
  {volume} {42}},\ \bibinfo {pages} {3--6} (\bibinfo {year}
  {1979})}\BibitemShut {NoStop}%
\bibitem [{Note3()}]{Note3}%
  \BibitemOpen
  \bibinfo {note} {The quantity $q$ plays the role of a conserved
  quasi-momentum in the single-particle QKR without periodic boundary
  conditions.}\BibitemShut {Stop}%
\bibitem [{\citenamefont {Vignolo}\ and\ \citenamefont
  {Minguzzi}(2013)}]{Vignolo2013}%
  \BibitemOpen
  \bibfield  {author} {\bibinfo {author} {\bibfnamefont {Patrizia}\
  \bibnamefont {Vignolo}}\ and\ \bibinfo {author} {\bibfnamefont {Anna}\
  \bibnamefont {Minguzzi}},\ }\bibfield  {title} {\enquote {\bibinfo {title}
  {Universal contact for a tonks-girardeau gas at finite temperature},}\ }\href
  {\doibase 10.1103/PhysRevLett.110.020403} {\bibfield  {journal} {\bibinfo
  {journal} {Phys. Rev. Lett.}\ }\textbf {\bibinfo {volume} {110}},\ \bibinfo
  {pages} {020403} (\bibinfo {year} {2013})}\BibitemShut {NoStop}%
\bibitem [{\citenamefont {Its}\ \emph {et~al.}(1991)\citenamefont {Its},
  \citenamefont {Izergin},\ and\ \citenamefont {Korepin}}]{Its1991}%
  \BibitemOpen
  \bibfield  {author} {\bibinfo {author} {\bibfnamefont {A.R.}\ \bibnamefont
  {Its}}, \bibinfo {author} {\bibfnamefont {A.G.}\ \bibnamefont {Izergin}}, \
  and\ \bibinfo {author} {\bibfnamefont {V.E.}\ \bibnamefont {Korepin}},\
  }\bibfield  {title} {\enquote {\bibinfo {title} {Space correlations in the
  one-dimensional impenetrable bose gas at finite temperature},}\ }\href
  {\doibase https://doi.org/10.1016/0167-2789(91)90171-5} {\bibfield  {journal}
  {\bibinfo  {journal} {Physica D: Nonlinear Phenomena}\ }\textbf {\bibinfo
  {volume} {53}},\ \bibinfo {pages} {187 -- 213} (\bibinfo {year}
  {1991})}\BibitemShut {NoStop}%
\bibitem [{\citenamefont {Collura}\ \emph {et~al.}(2013)\citenamefont
  {Collura}, \citenamefont {Sotiriadis},\ and\ \citenamefont
  {Calabrese}}]{Collura2013}%
  \BibitemOpen
  \bibfield  {author} {\bibinfo {author} {\bibfnamefont {Mario}\ \bibnamefont
  {Collura}}, \bibinfo {author} {\bibfnamefont {Spyros}\ \bibnamefont
  {Sotiriadis}}, \ and\ \bibinfo {author} {\bibfnamefont {Pasquale}\
  \bibnamefont {Calabrese}},\ }\bibfield  {title} {\enquote {\bibinfo {title}
  {Equilibration of a tonks-girardeau gas following a trap release},}\ }\href
  {\doibase 10.1103/PhysRevLett.110.245301} {\bibfield  {journal} {\bibinfo
  {journal} {Phys. Rev. Lett.}\ }\textbf {\bibinfo {volume} {110}},\ \bibinfo
  {pages} {245301} (\bibinfo {year} {2013})}\BibitemShut {NoStop}%
\bibitem [{Note4()}]{Note4}%
  \BibitemOpen
  \bibinfo {note} {The Periodic Gibbs ensemble allows for a time dependence of
  the steady-state density matrix $\rho _{GGE}(t)$ with $0\leq t<1$, with
  period $1$. We only consider here the case $t=1^-$ (just before the next
  kick). We have checked that other choices do not change the qualitative
  picture.}\BibitemShut {Stop}%
\bibitem [{Vua()}]{Vuatelet_tobepublished}%
  \BibitemOpen
  \href@noop {} {}\bibinfo {note} {V. Vuatelet and A. Ran\c{c}on, to be
  published.}\BibitemShut {Stop}%
\bibitem [{\citenamefont {van Amerongen}\ \emph {et~al.}(2008)\citenamefont
  {van Amerongen}, \citenamefont {van Es}, \citenamefont {Wicke}, \citenamefont
  {Kheruntsyan},\ and\ \citenamefont {van Druten}}]{Amerongen2008}%
  \BibitemOpen
  \bibfield  {author} {\bibinfo {author} {\bibfnamefont {A.~H.}\ \bibnamefont
  {van Amerongen}}, \bibinfo {author} {\bibfnamefont {J.~J.~P.}\ \bibnamefont
  {van Es}}, \bibinfo {author} {\bibfnamefont {P.}~\bibnamefont {Wicke}},
  \bibinfo {author} {\bibfnamefont {K.~V.}\ \bibnamefont {Kheruntsyan}}, \ and\
  \bibinfo {author} {\bibfnamefont {N.~J.}\ \bibnamefont {van Druten}},\
  }\bibfield  {title} {\enquote {\bibinfo {title} {Yang-yang thermodynamics on
  an atom chip},}\ }\href {\doibase 10.1103/PhysRevLett.100.090402} {\bibfield
  {journal} {\bibinfo  {journal} {Phys. Rev. Lett.}\ }\textbf {\bibinfo
  {volume} {100}},\ \bibinfo {pages} {090402} (\bibinfo {year}
  {2008})}\BibitemShut {NoStop}%
\bibitem [{\citenamefont {Fabbri}\ \emph {et~al.}(2011)\citenamefont {Fabbri},
  \citenamefont {Cl\'ement}, \citenamefont {Fallani}, \citenamefont {Fort},\
  and\ \citenamefont {Inguscio}}]{Fabbri2011}%
  \BibitemOpen
  \bibfield  {author} {\bibinfo {author} {\bibfnamefont {N.}~\bibnamefont
  {Fabbri}}, \bibinfo {author} {\bibfnamefont {D.}~\bibnamefont {Cl\'ement}},
  \bibinfo {author} {\bibfnamefont {L.}~\bibnamefont {Fallani}}, \bibinfo
  {author} {\bibfnamefont {C.}~\bibnamefont {Fort}}, \ and\ \bibinfo {author}
  {\bibfnamefont {M.}~\bibnamefont {Inguscio}},\ }\bibfield  {title} {\enquote
  {\bibinfo {title} {Momentum-resolved study of an array of one-dimensional
  strongly phase-fluctuating bose gases},}\ }\href {\doibase
  10.1103/PhysRevA.83.031604} {\bibfield  {journal} {\bibinfo  {journal} {Phys.
  Rev. A}\ }\textbf {\bibinfo {volume} {83}},\ \bibinfo {pages} {031604}
  (\bibinfo {year} {2011})}\BibitemShut {NoStop}%
\bibitem [{Note5()}]{Note5}%
  \BibitemOpen
  \bibinfo {note} {Using $T_{eff}\simeq \protect \sqrt {\varepsilon
  _{loc}\varepsilon _{F}}$, with $\varepsilon _{loc}$ the localization energy
  per particles, and the estimates $\varepsilon _F\simeq E_R N^2$ and
  $\varepsilon _{loc}\simeq \alpha E_R$ with $E_R$ the recoil energy and
  $\alpha $ typically of order a few hundreds depending on the kicks
  parameters, we estimate that $T_{eff}\simeq 10 N T_R$ ($T_R=E_R/k_B$ the
  recoil temperature, typically of order $ 10^{-7} K$ in cold atoms
  experiments).}\BibitemShut {Stop}%
\bibitem [{\citenamefont {Wilson}\ \emph {et~al.}(2020)\citenamefont {Wilson},
  \citenamefont {Malvania}, \citenamefont {Le}, \citenamefont {Zhang},
  \citenamefont {Rigol},\ and\ \citenamefont {Weiss}}]{Wilson2020}%
  \BibitemOpen
  \bibfield  {author} {\bibinfo {author} {\bibfnamefont {Joshua~M.}\
  \bibnamefont {Wilson}}, \bibinfo {author} {\bibfnamefont {Neel}\ \bibnamefont
  {Malvania}}, \bibinfo {author} {\bibfnamefont {Yuan}\ \bibnamefont {Le}},
  \bibinfo {author} {\bibfnamefont {Yicheng}\ \bibnamefont {Zhang}}, \bibinfo
  {author} {\bibfnamefont {Marcos}\ \bibnamefont {Rigol}}, \ and\ \bibinfo
  {author} {\bibfnamefont {David~S.}\ \bibnamefont {Weiss}},\ }\bibfield
  {title} {\enquote {\bibinfo {title} {Observation of dynamical
  fermionization},}\ }\href {\doibase 10.1126/science.aaz0242} {\bibfield
  {journal} {\bibinfo  {journal} {Science}\ }\textbf {\bibinfo {volume}
  {367}},\ \bibinfo {pages} {1461--1464} (\bibinfo {year} {2020})}\BibitemShut
  {NoStop}%
\bibitem [{\citenamefont {Malvania}\ \emph {et~al.}(2021)\citenamefont
  {Malvania}, \citenamefont {Zhang}, \citenamefont {Le}, \citenamefont
  {Dubail}, \citenamefont {Rigol},\ and\ \citenamefont {Weiss}}]{Malvania2021}%
  \BibitemOpen
  \bibfield  {author} {\bibinfo {author} {\bibfnamefont {Neel}\ \bibnamefont
  {Malvania}}, \bibinfo {author} {\bibfnamefont {Yicheng}\ \bibnamefont
  {Zhang}}, \bibinfo {author} {\bibfnamefont {Yuan}\ \bibnamefont {Le}},
  \bibinfo {author} {\bibfnamefont {Jerome}\ \bibnamefont {Dubail}}, \bibinfo
  {author} {\bibfnamefont {Marcos}\ \bibnamefont {Rigol}}, \ and\ \bibinfo
  {author} {\bibfnamefont {David~S.}\ \bibnamefont {Weiss}},\ }\bibfield
  {title} {\enquote {\bibinfo {title} {Generalized hydrodynamics in strongly
  interacting 1d bose gases},}\ }\href {\doibase 10.1126/science.abf0147}
  {\bibfield  {journal} {\bibinfo  {journal} {Science}\ }\textbf {\bibinfo
  {volume} {373}},\ \bibinfo {pages} {1129--1133} (\bibinfo {year}
  {2021})}\BibitemShut {NoStop}%
\bibitem [{\citenamefont {Shepelyansky}(1987)}]{Shepelyansky1987}%
  \BibitemOpen
  \bibfield  {author} {\bibinfo {author} {\bibfnamefont {D.L.}\ \bibnamefont
  {Shepelyansky}},\ }\bibfield  {title} {\enquote {\bibinfo {title}
  {Localization of diffusive excitation in multi-level systems},}\ }\href
  {\doibase https://doi.org/10.1016/0167-2789(87)90123-0} {\bibfield  {journal}
  {\bibinfo  {journal} {Physica D: Nonlinear Phenomena}\ }\textbf {\bibinfo
  {volume} {28}},\ \bibinfo {pages} {103 -- 114} (\bibinfo {year}
  {1987})}\BibitemShut {NoStop}%
\bibitem [{\citenamefont {Casati}\ \emph {et~al.}(1989)\citenamefont {Casati},
  \citenamefont {Guarneri},\ and\ \citenamefont {Shepelyansky}}]{Casati1989}%
  \BibitemOpen
  \bibfield  {author} {\bibinfo {author} {\bibfnamefont {Giulio}\ \bibnamefont
  {Casati}}, \bibinfo {author} {\bibfnamefont {Italo}\ \bibnamefont
  {Guarneri}}, \ and\ \bibinfo {author} {\bibfnamefont {D.~L.}\ \bibnamefont
  {Shepelyansky}},\ }\bibfield  {title} {\enquote {\bibinfo {title} {Anderson
  transition in a one-dimensional system with three incommensurate
  frequencies},}\ }\href {\doibase 10.1103/PhysRevLett.62.345} {\bibfield
  {journal} {\bibinfo  {journal} {Phys. Rev. Lett.}\ }\textbf {\bibinfo
  {volume} {62}},\ \bibinfo {pages} {345--348} (\bibinfo {year}
  {1989})}\BibitemShut {NoStop}%
\bibitem [{\citenamefont {Radi{\'c}}\ \emph {et~al.}(2010)\citenamefont
  {Radi{\'c}}, \citenamefont {Ba{\v c}i{\'c}}, \citenamefont {Juki{\'c}},
  \citenamefont {Segev},\ and\ \citenamefont {Buljan}}]{Radic2010}%
  \BibitemOpen
  \bibfield  {author} {\bibinfo {author} {\bibfnamefont {J.}~\bibnamefont
  {Radi{\'c}}}, \bibinfo {author} {\bibfnamefont {V.}~\bibnamefont {Ba{\v
  c}i{\'c}}}, \bibinfo {author} {\bibfnamefont {D.}~\bibnamefont {Juki{\'c}}},
  \bibinfo {author} {\bibfnamefont {M.}~\bibnamefont {Segev}}, \ and\ \bibinfo
  {author} {\bibfnamefont {H.}~\bibnamefont {Buljan}},\ }\bibfield  {title}
  {\enquote {\bibinfo {title} {Anderson localization of a tonks-girardeau gas
  in potentials with controlled disorder},}\ }\href {\doibase
  10.1103/PhysRevA.81.063639} {\bibfield  {journal} {\bibinfo  {journal}
  {Physical Review A}\ }\textbf {\bibinfo {volume} {81}},\ \bibinfo {pages}
  {063639--} (\bibinfo {year} {2010})}\BibitemShut {NoStop}%
\bibitem [{\citenamefont {Seiringer}\ and\ \citenamefont
  {Warzel}(2016)}]{Seiringer2016}%
  \BibitemOpen
  \bibfield  {author} {\bibinfo {author} {\bibfnamefont {Robert}\ \bibnamefont
  {Seiringer}}\ and\ \bibinfo {author} {\bibfnamefont {Simone}\ \bibnamefont
  {Warzel}},\ }\bibfield  {title} {\enquote {\bibinfo {title} {Decay of
  correlations and absence of superfluidity in the disordered tonks--girardeau
  gas},}\ }\href {\doibase 10.1088/1367-2630/18/3/035002} {\bibfield  {journal}
  {\bibinfo  {journal} {New Journal of Physics}\ }\textbf {\bibinfo {volume}
  {18}},\ \bibinfo {pages} {035002} (\bibinfo {year} {2016})}\BibitemShut
  {NoStop}%
\end{thebibliography}%

\appendix
\section{Validity of the thermal fit of the steady-state of the fermionic momentum distributions \label{app_fit}}

We focus here on giving details on the effective thermalization and its range of validity with respect to the parameters of the system.

\begin{figure}[t]
	\includegraphics[scale=0.4,clip]{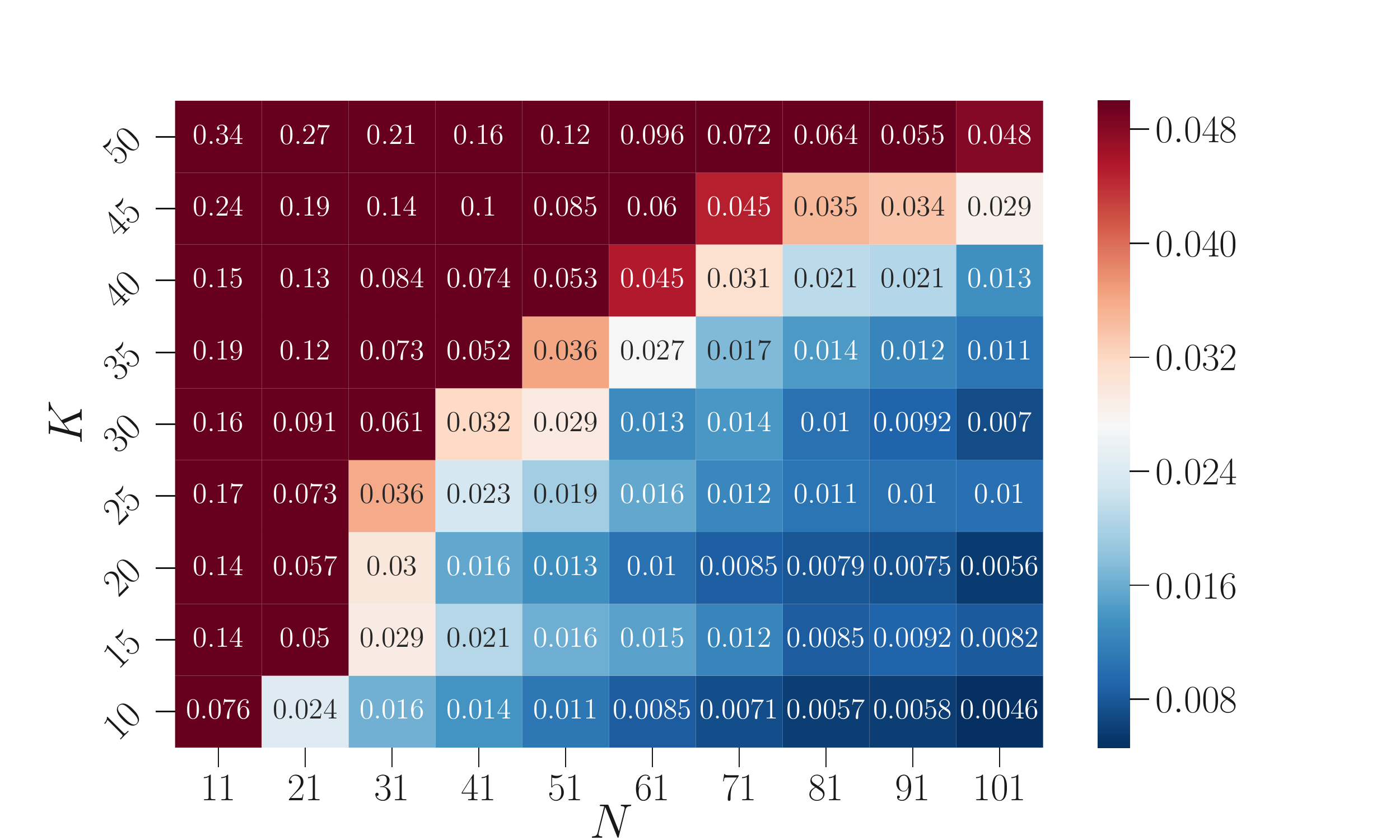}
	\includegraphics[scale=0.4,clip]{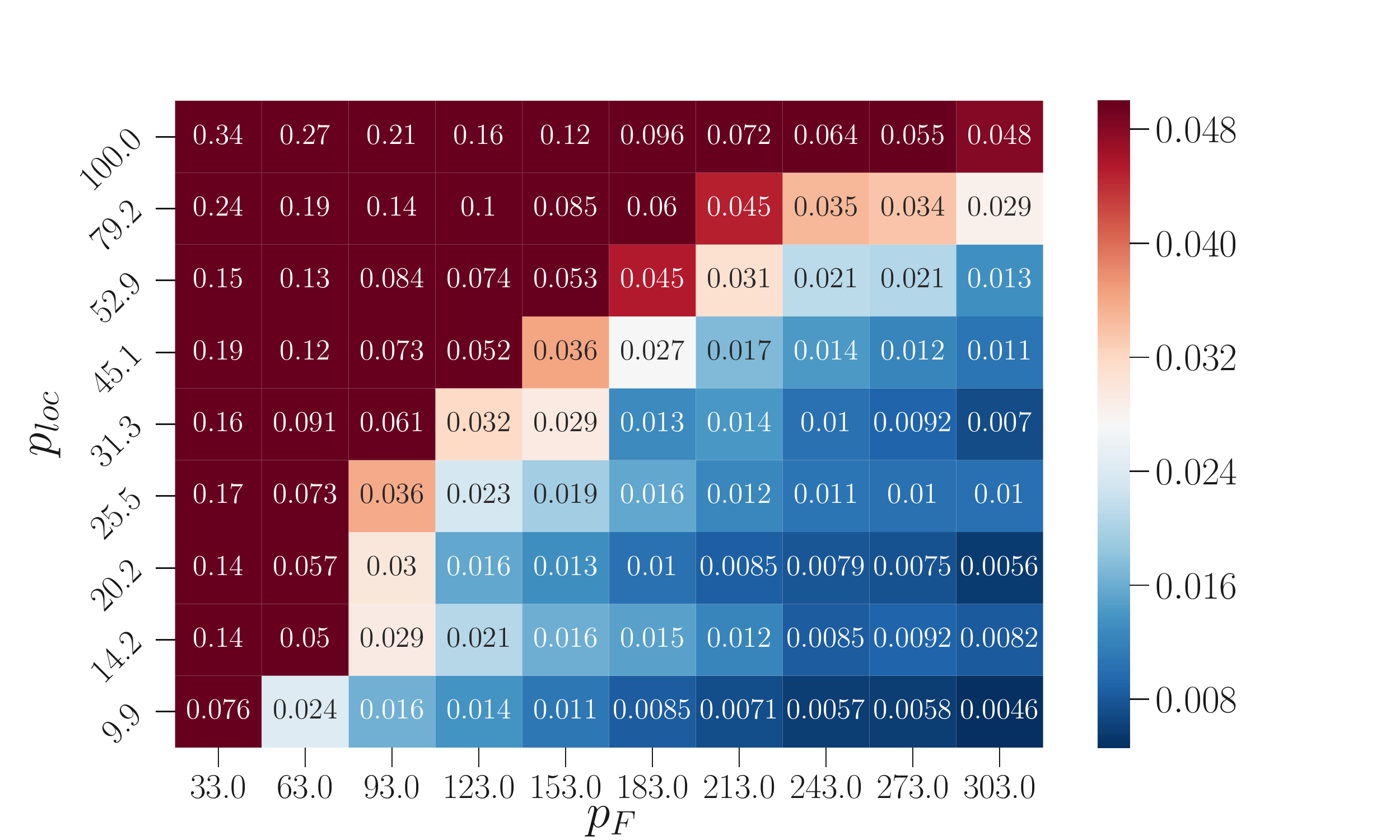}
	\caption{Top panel: $\varepsilon(N,K)$ at fixed $\kbar=6$. Dark red color corresponds to $\varepsilon\geq0.05$. Bottom panel: same as top panel but in function of $p_{loc}$ and $p_{F}$.}
	\label{fig_epsilon_pF_ploc}
\end{figure}

As hinted in the inset of Fig.~\ref{fig_fermion_thermal}, the thermal fit of the steady-state  fermionic momentum distributions works well for $p_{loc}/p_{F}\ll 1$, while in the opposite limit, it does not work, implying that the system does not effectively thermalize. This can be quantified by introducing the error 
\begin{equation}
\varepsilon = \frac{\lVert n^{F}_{k}-f_{FD}(k,T_{eff},\mu_{eff})\rVert}{\lVert n^{F}_{k}\rVert}.
\end{equation}
The error as a function of $N$ and $K$ is shown in the top panel of Fig.~\ref{fig_epsilon_pF_ploc}, and as a function of $p_F$ and $p_{loc}$ in the bottom panel. The low value of $\varepsilon$, i.e. a good the thermal fit, corresponds to the  blue area in Fig.~\ref{fig_epsilon_pF_ploc}). In the following, we only consider parameters such that $\varepsilon\lesssim 5\%$, where the effective thermalization takes place. 

\begin{figure}[t!]
	\begin{center}
		\includegraphics[scale=0.2]{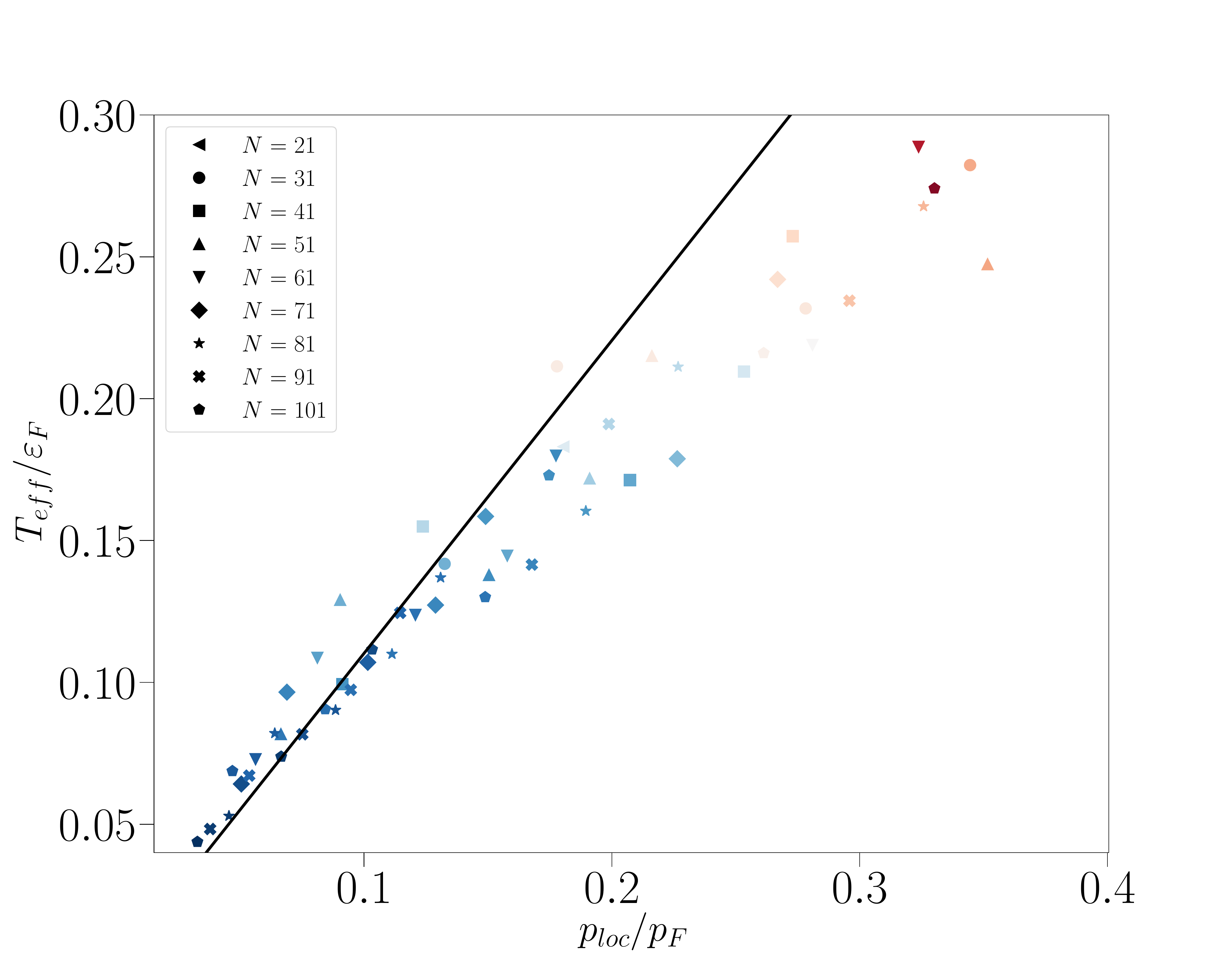}
		\caption{Effective temperature $T_{eff}/\varepsilon_{F}$ as a function of $p_{loc}/p_{F}$ for $\kb=6$ and various particle number. In practice, we plot $\frac{E_{f}-E_{0}}{N\varepsilon_{F}}$ as a function of $T_{eff}^{2}/\varepsilon_{F}^{2}$. The black line corresponds to the prediction Eq.~\eqref{eq_Teff_ploc}. The color code corresponds to that of Fig.~\ref{fig_epsilon_pF_ploc}.}
		\label{figteffeF_vs_plocpF}
	\end{center}
\end{figure}

Fig.~\ref{figure rescaled	energy} shows that the localized energy in the steady-state is well described by the Sommerfeld expansion Eq.~\eqref{eq_energy} in terms of the fitted effective temperature $T_{eff}$, as long as the temperature is small enough, i.e. when the thermal fit works well (dark blue symbols). Fig.~\ref{figteffeF_vs_plocpF} shows that in the same regime, the effective temperature is also well described by our prediction Eq.~ \eqref{eq_Teff_ploc}.

\begin{figure}[t!]
	\includegraphics[scale=0.2,clip]{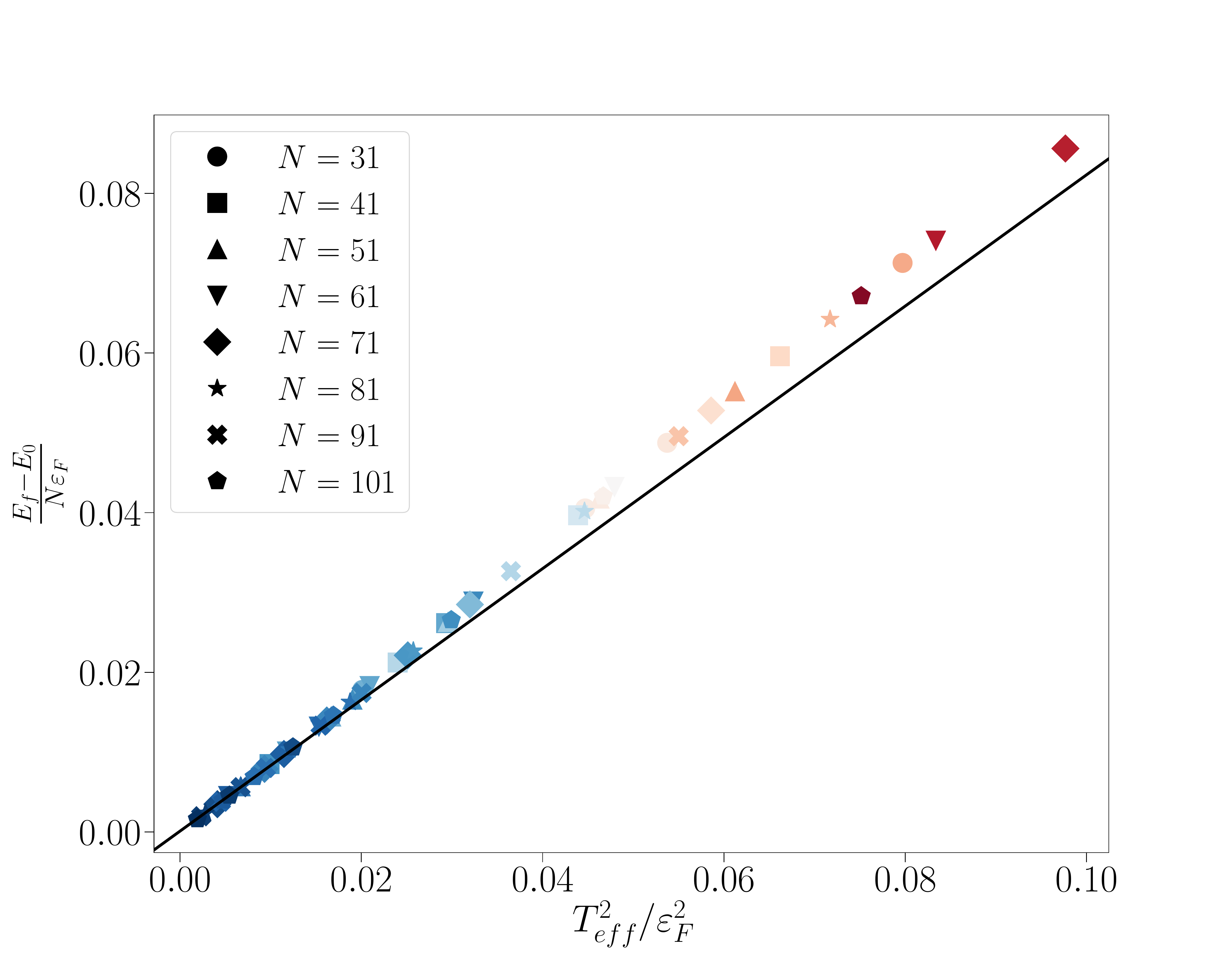}
	\caption{Final energy $E_f$ as a function of the fitted effective temperature $T_{eff}$ for $\kb=6$ and various particle number. In practice, we plot $\frac{E_{f}-E_{0}}{N\varepsilon_{F}}$ as a function of $T_{eff}^{2}/\varepsilon_{F}^{2}$. The black line corresponds to the Sommerfeld expansion in Eq.~\eqref{eq_energy}. The color code corresponds to that of Fig.~\ref{fig_epsilon_pF_ploc}.} 
	\label{figure rescaled energy}
\end{figure}

\begin{figure}[h]
	\includegraphics[scale=0.2,clip]{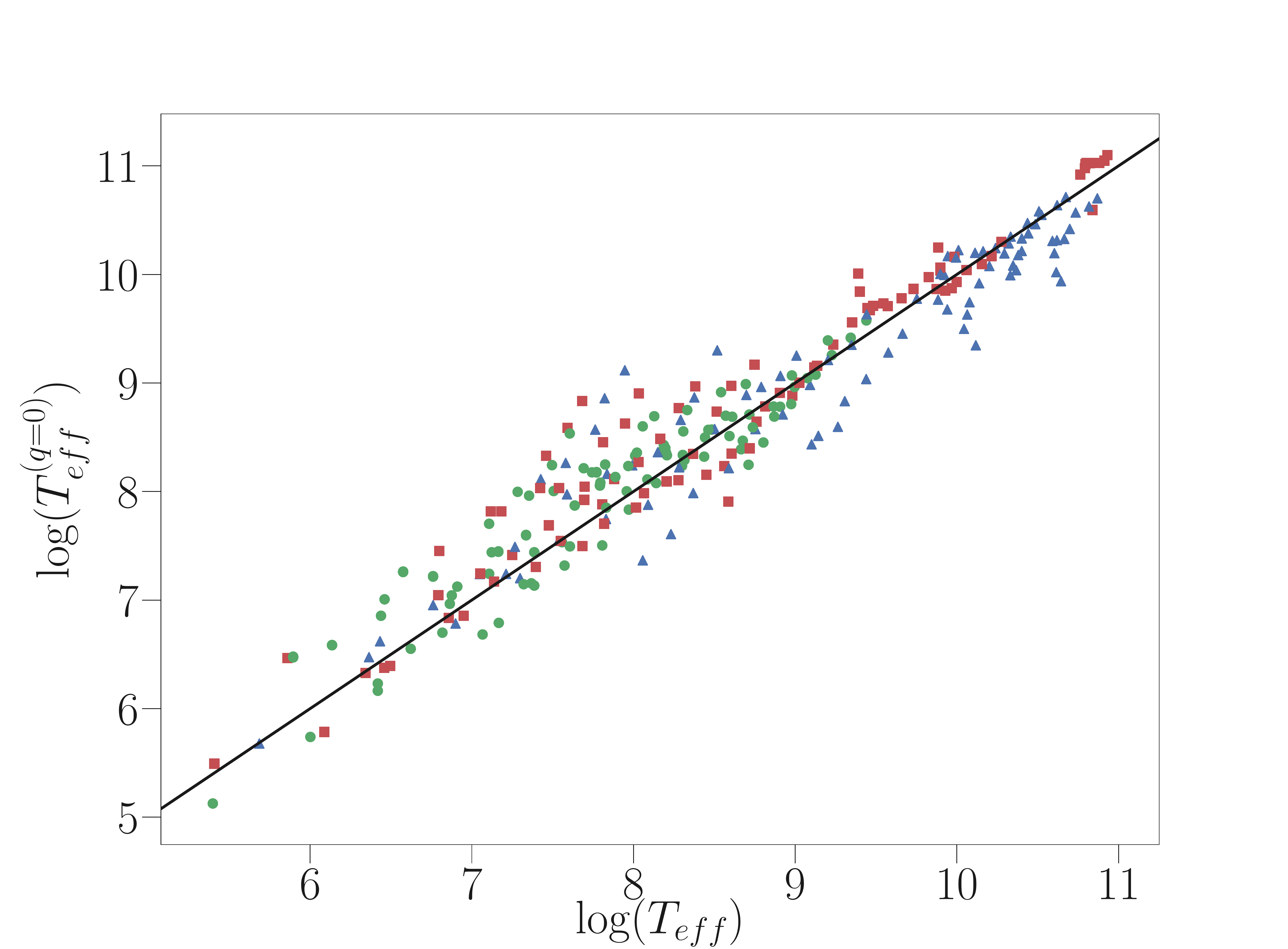}
	\caption{Comparaison of temperature extracted from $E_{f}$ and from the averaged distribution (respectively $T_{eff}^{(q=0)}$ and $T_{eff}$). The colored dots are the datas (green: $\kbar=6$, red : $\kbar=7$, blue: $\kbar=8$) and the black line is a guide to the eye. The collapse of the datas around the black line show that $T_{eff}^{(q=0)}$ and $T_{eff}$ are mostly the same.}
	\label{fig_comp_avg}
\end{figure}

Finally, let us address the effects of the averaging over $q$. Fig.~\ref{fig_comp_avg} is a scatter plot of $T_{eff}^{(q=0)}$, the effective temperature extracted from the fermionic energy for $q=0$, and  $T_{eff}$, the effective temperature obtained from the averaged momentum distribution, for various values of $N$, $\kbar$ and $K$. We see a very clear correlation between the two. This shows that while averaging is convenient to analyze the fermionic degrees of freedom, the effective temperature and chemical potential obtained will describe very well the non-averaged observables of the bosons.

\newpage

\section{Extraction of $r_c$ \label{app_rc}}

\begin{figure}
	\begin{center}
		\includegraphics[scale=0.2,clip]{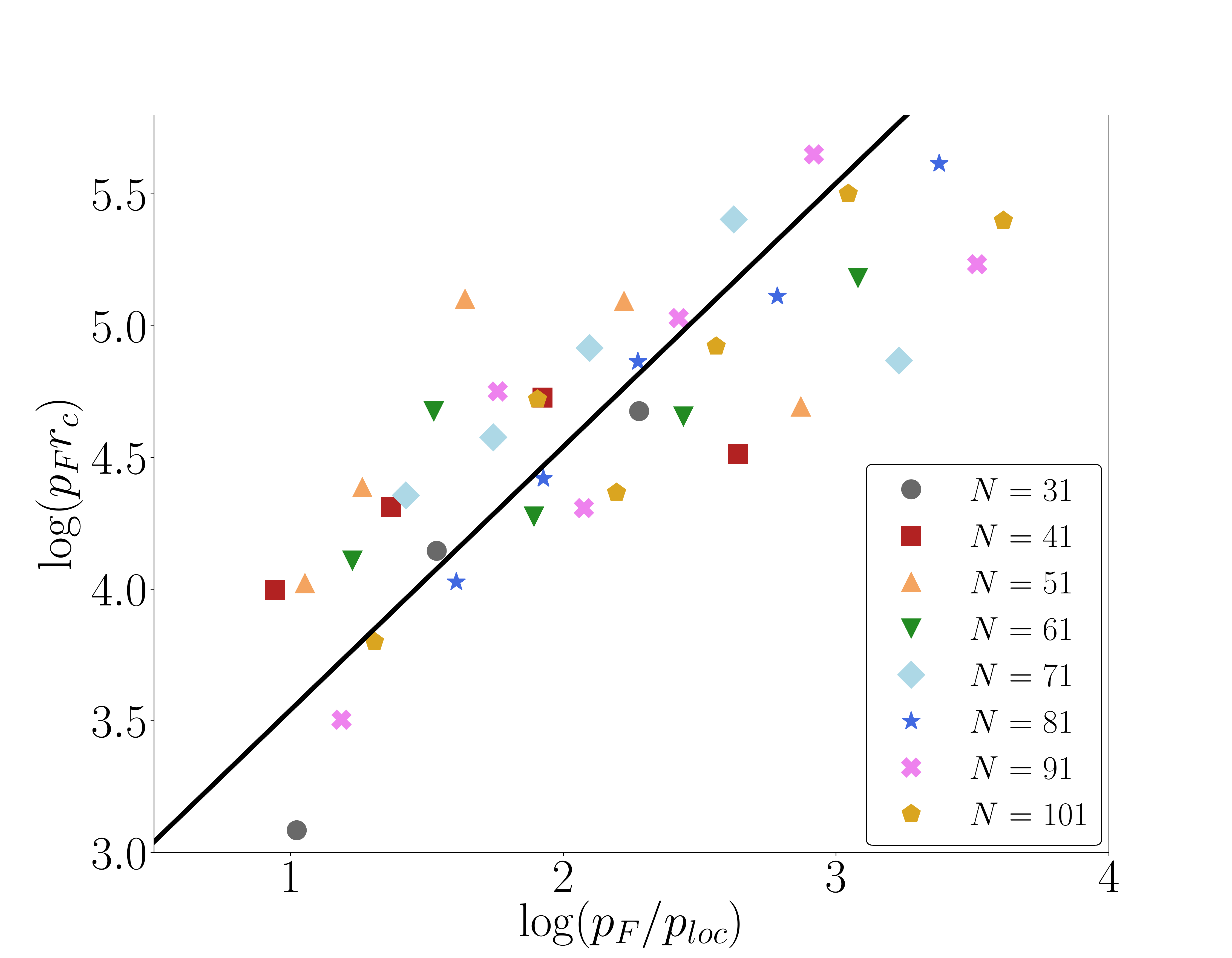}
		\includegraphics[scale=0.2,clip]{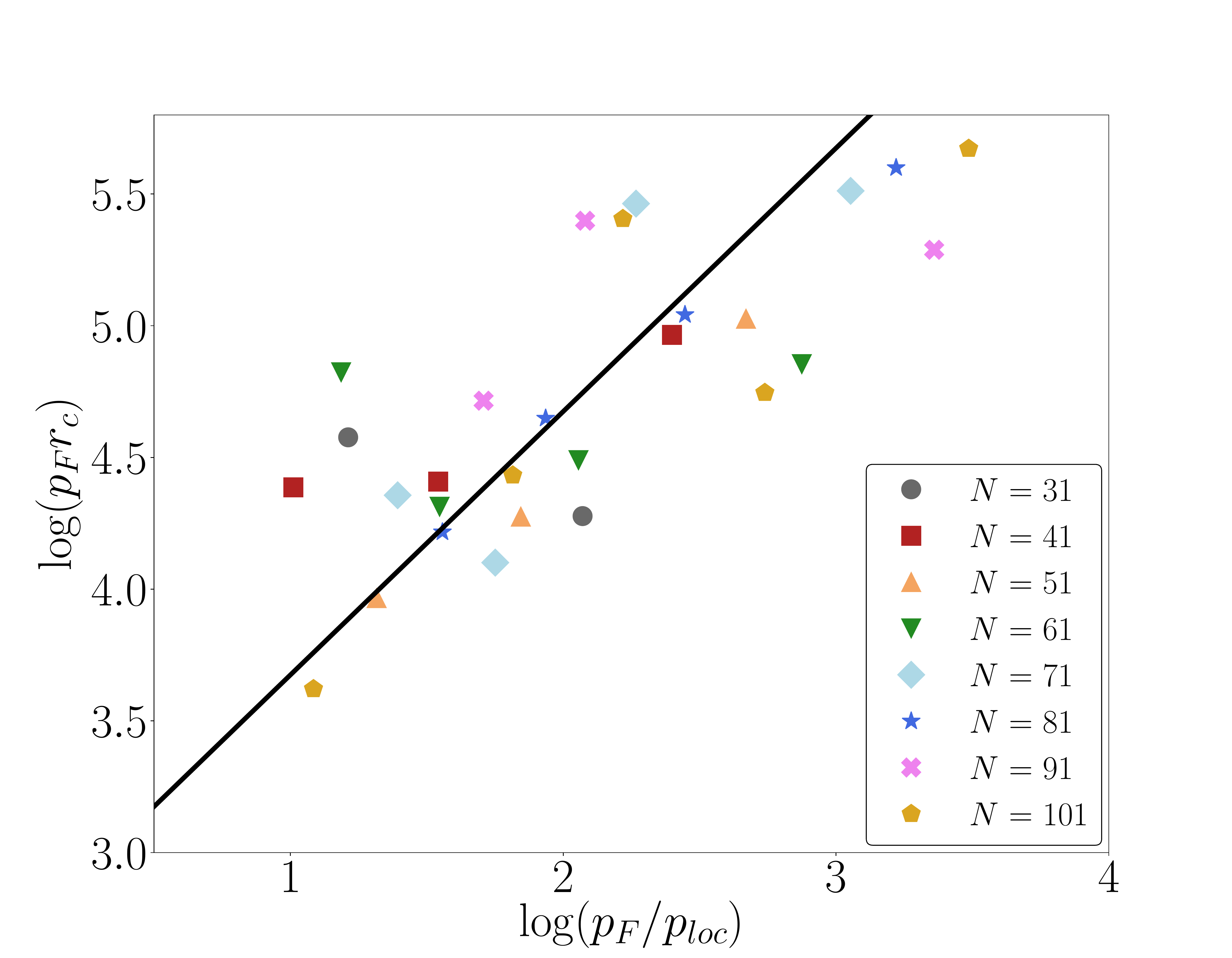}
	
		\caption{Correlation length for different $\kbar$ (7, 8 from up to down). The black line correspond on every figure to the relation Eq.~\eqref{eq_rc}. }
		\label{fig_rc_over_N_vs_1_over_teff_some_kbar}
	\end{center}
\end{figure}

We observed that the coherence function of the Tonks gas $g_1(r)$ decays exponentially in the localized regime as it was shown in Fig.~\ref{fig_g1}. Assuming that it decays as $g_1(r)\propto e^{-2|r|/r_c}$, we can estimate the correlation length $r_c$ by
\begin{equation}
r_{c}=\sqrt{2\frac{\sum_{r}r^{2}g_{1}(r)-(\sum_{r}rg_{1}(r))^{2}}{\sum_{r}g_{1}(r)}},
\end{equation}
which is well described by Eq.~\eqref{eq_rc} as discussed in the main text, where we have focus on the case $\kbar=6$. We show in Fig.~\ref{fig_rc_over_N_vs_1_over_teff_some_kbar} that for other $\kbar$, our prediction is in good agreement with the data.

\section{Natural orbitals \label{app_orbitals}}

\begin{figure}[h!!!!]
\begin{center}	\includegraphics[scale=0.22,clip]{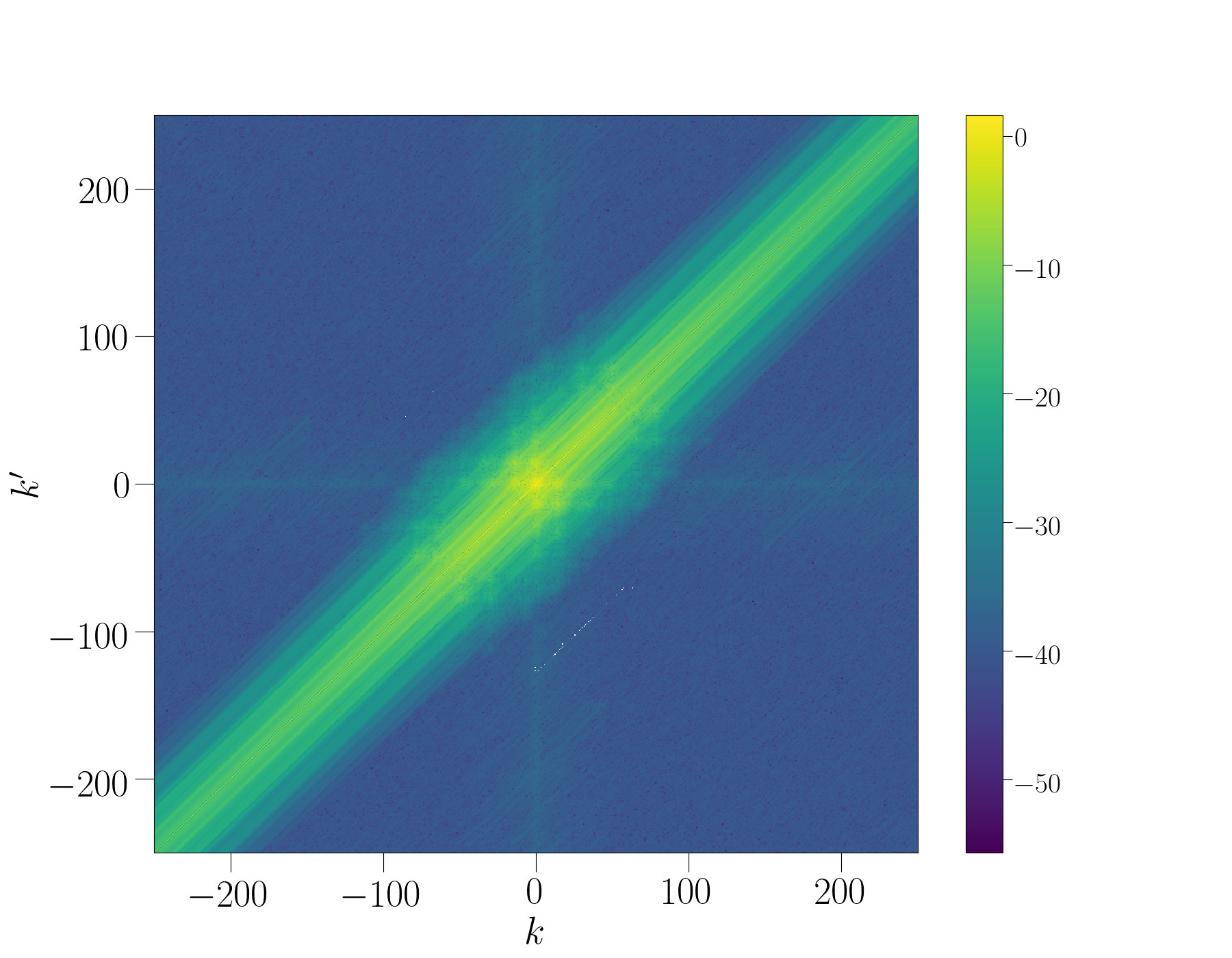}
	
\includegraphics[scale=0.2,clip]{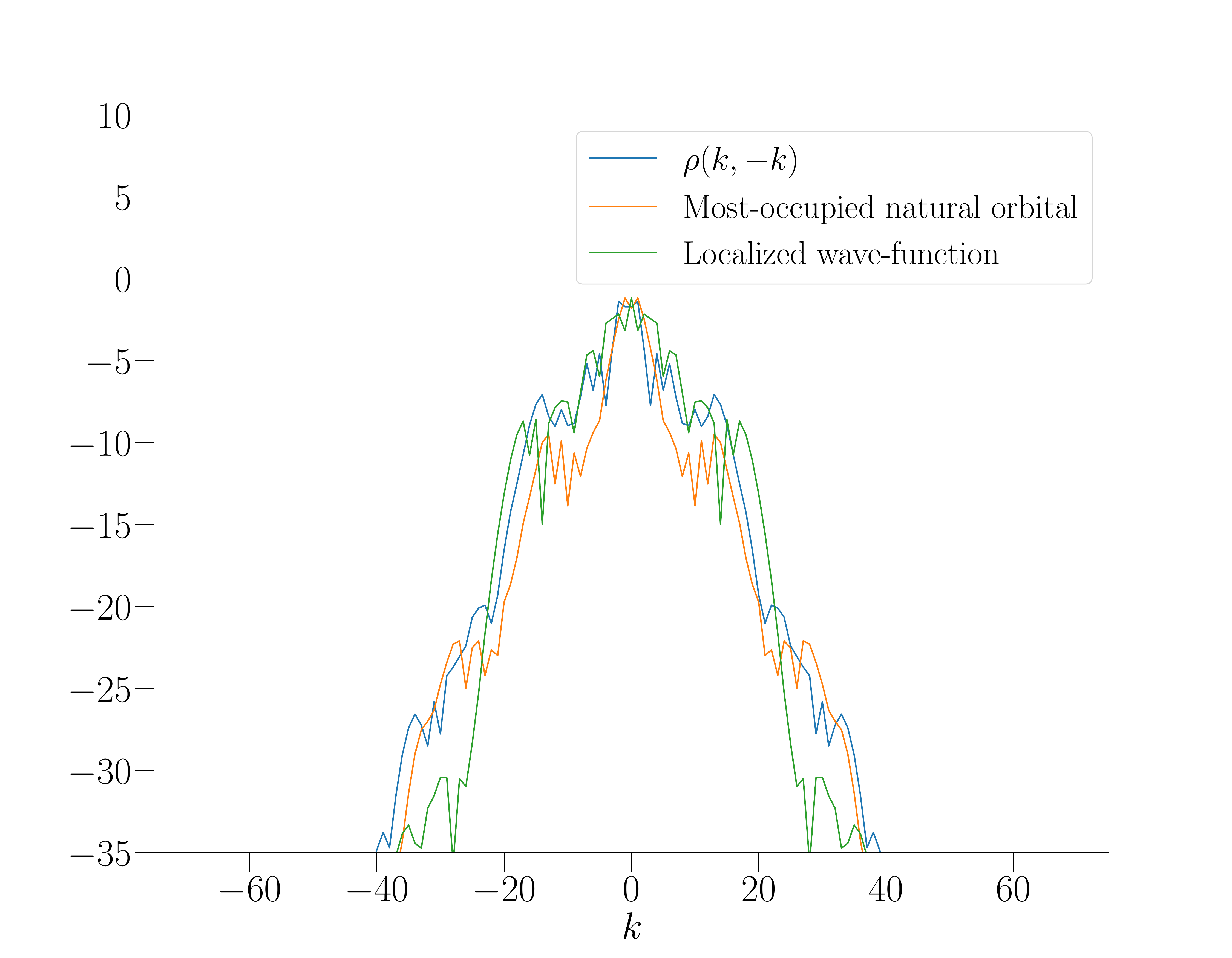}

\end{center}
\caption{In both panel $N=51$, $K=20$, $\kbar=6$. Top panel: 2D graphic of the logarithm value of the OBDM in momentum space $\rho(k,k')$. Bottom panel: comparison between the anti-diagonal of the OBDM in momentum space and the most-occupied natural orbitals. We compare them to the $k=0$ wave-function for the same parameters.}	
\label{figorbitalnaturelles}
\end{figure}

The OBDM can be decomposed in natural orbitals $\phi^{\eta}(x)$, which can be interpreted as the many-body version of the wavefunctions occupied by the bosons, and which are the eigenfunctions of the OBDM, 
\begin{equation}
\int dy \rho(x,y)\phi^{\eta}(y)=\lambda_{\eta}\phi^{\eta}(x),
\end{equation}
with the $\lambda_{\eta}$the occupation of $\eta$-th natural orbital.  Fig.~\ref{figorbitalnaturelles} (top) shows the most occupied natural orbital in momentum space for $N=51$, $K=20$, $\kbar=6$ in semi-log scale. We observe that it decays exponentially over a scale $p_{loc}$, as can been verified by plotting a localized wave-function of the non-interacting QKR (which decays over the same scale). 

Fig.~\ref{figorbitalnaturelles} (top) shows the two-dimensional Fourier transform of the OBDM,
\begin{equation}
\rho(k,k')=\frac{1}{L}\int dx dy e^{ikx-ik'y}\rho(x,y),
\end{equation}
where $\rho(k,k)$ is the momentum distribution. We observe that contrary to a thermal OBDM, it is non-zero for $k\neq k'$ (as expected by invariance by translation for the thermal gas). However, it decays exponentially over the scale $p_{loc}$, as can be seen in Fig.~\ref{figorbitalnaturelles} (bottom).

\end{document}